\documentclass{aa}
\usepackage[varg]{txfonts}
\usepackage{graphicx}
\usepackage{times}
\usepackage{color}

\definecolor{customhdrcolor}{rgb}{0.0,0.0,0.0}
\definecolor{customcitecolor}{rgb}{0.0,0.5,0.75}
\definecolor{customlinkcolor}{rgb}{0.0,0.5,0.75}

\usepackage[colorlinks=true,linkcolor=customlinkcolor,urlcolor=customlinkcolor,citecolor=customcitecolor,pdftex]{hyperref}
\usepackage[caption=false]{subfig}

\ifpdf\else\usepackage{graphics}\fi

\DeclareRobustCommand{\TUSSEN}[3]{#2}

\begin{document}

   \title{Compression of interferometric radio-astronomical data}


   \author{A. R. Offringa}          

   \institute{Netherlands Institute for Radio Astronomy (ASTRON), PO Box 2, 7990 AA Dwingeloo, The Netherlands\\ \email{offringa@astron.nl}}

   \date{Received August 22, 2016 / Accepted September 7, 2016}

\label{firstpage}
  \abstract
   {The volume of radio-astronomical data is a considerable burden in the processing and storing of radio observations with high time and frequency resolutions and large bandwidths. For future telescopes such as the SKA, the data volume will be even larger.}
   {Lossy compression of interferometric radio-astronomical data is considered to reduce the volume of visibility data and to speed up processing.}
   {A new compression technique named ``Dysco'' is introduced that consists of two steps: a normalization step, in which grouped visibilities are
   normalized to have a similar distribution; and a quantization and encoding step, which rounds values to a given quantization scheme using a dithering scheme. Several non-linear quantization schemes are tested and combined with different methods for normalizing the data. Four data sets with observations from the LOFAR and MWA telescopes are processed with different processing strategies and different combinations of normalization and quantization. The effects of compression are measured in image plane.}
   {The noise added by the lossy compression technique acts like normal system noise. The accuracy of Dysco is depending on the signal-to-noise ratio of the data: noisy data can be compressed with a smaller loss of image quality. Data with typical correlator time and frequency resolutions can be compressed by a factor of 6.4 for LOFAR and 5.3 for MWA observations with less than 1\% added system noise. An implementation of the compression technique is released that provides a Casacore storage manager and allows transparent encoding and decoding. Encoding and decoding is faster than the read/write speed of typical disks.}
   {The technique can be used for LOFAR and MWA to reduce the archival space requirements for storing observed data. Data from SKA-low will likely be compressible by the same amount as LOFAR. The same technique can be used to compress data from other telescopes, but a different bit-rate might be required.}


   \maketitle
   
   \titlerunning{Compression of interferometric radio-astronomical data}
   \authorrunning{A. R. Offringa}
%

\section{Background}
Data from interferometric radio observatories consists of correlated signals from all the pairs of antennae that are part of the interferometric array. These correlated signals are measured in spectral channels, and integrated over short timesteps before they are recorded to disk \citep{thompson-radio-interferometry}.

Correlators in modern interferometric radio observatories output data at high spectral and temporal resolutions \citep{evla-perley-2011,atca-broadband-backend-2011,mwa-2013-tingay,lofar-2013}. It is often not desirable to decrease the resolution before archiving, because resolution in both dimensions provides scientific value. High temporal resolution is for example desirable for variability studies~\citep{lofar-pulsars-and-transients-2011}, advanced calibration methods that solve for the ionospheric disturbances and instrumental variabilities \citep{kazemi-clustered-cal-2013,revisiting-me-ii,vanweeren-factor-2016} and to avoid time decorrelation. On the other hand, high frequency resolution is necessary for spectral-line studies~\citep{evla-high-resolution-lines,morabito-2014-carbon-rrls,offringa-2016}, accurate removal of man-made interference~\citep{lofar-radio-environment} and bandwidth decorrelation. The required time and frequency resolutions can lead to data volumes of several petabytes, which have to be supported by complex network architectures to allow acceptable transfer rates. While current interferometers consist of up to a few hundred of antenna elements, future instruments will have thousands of antenna elements that will produce data at even higher resolutions or with multiple beams, thereby increasing the data volume and associated storage costs by several orders of magnitudes \citep{apertif-2010,ska-station-config-2013,heywood-2016-askap}. For these reasons, compression of correlated interferometric data is desirable.

In radio astronomy, linear quantization of the captured antenna voltages \textit{prior} to correlation is common and well understood \citep{quantization-efficiency-2007}. Further quantization of correlated data is however not common, mainly because it has not been explored what the effects of quantization are on the final science products, such as images or spectra. Since many interferometric projects need to reach large dynamic ranges, compression techniques that would limit the dynamic range are undesirable.

This paper considers compressing correlator output samples that consist of complex values. These data commonly have a low signal-to-noise ratio (SNR), especially at high resolutions \citep{thompson-radio-interferometry}, and noise-like signals are inherently hard to compress without loss of information \citep{shannon-entropy-definition-1949}. In radio data, each complex component is normally stored as a 32-bit IEEE 754 single-precision floating-point value\footnote{See e.g. ``FITS Interferometry Data Interchange Format'', AIPS memo \#114, \S4.1; and ``MeasurementSet definition version 2.0'', CASA Memo \#229.}. Loss-less compression is only effective for the 8-bit exponents of the floating point values, because the mantissa and sign follow closely a uniform distribution and are incompressible without loss. Therefore, loss-less compression can only reduce the volume of visibilities to approximately 75\% of the original size. For these reasons, this paper will explore compression with a lossy encoding.

There are a few existing implementations of compression that are related to compressing visibilities:
\begin{itemize}
 \item \textsc{aips} can write visibilities as 16-bit values with uniform quantization;
 \item The Flexible Image Transport System (FITS) file format \citep{1982-wells-fitsformat} allows Rice, GZIP, HCompress, and PLIO compression\footnote{See \url{https://heasarc.gsfc.nasa.gov/docs/software/fitsio/compression.html}.}.
 \item For compression of integer data recorded with the CHIME telescope, \citet{masui-2015-compression} showed that those data can be compressed to 28\% of its original size using linear quantization and applying the \textsc{bitshuffle} technique before LZ77 encoding \citep{ziv-1977-lz77}.
\end{itemize}
FITS compression was designed particularly with image compression in mind, and is not designed for visibility data. FITS compression uses exponential quantization (the mantissa is rounded). It supports subtractive dithering \citep{pence-2010-fits-dithering}. \textsc{aips} and \textsc{bitshuffle} do not use dithering.

\begin{figure*}
\centering
\subfloat[Original]{\includegraphics[height=7cm]{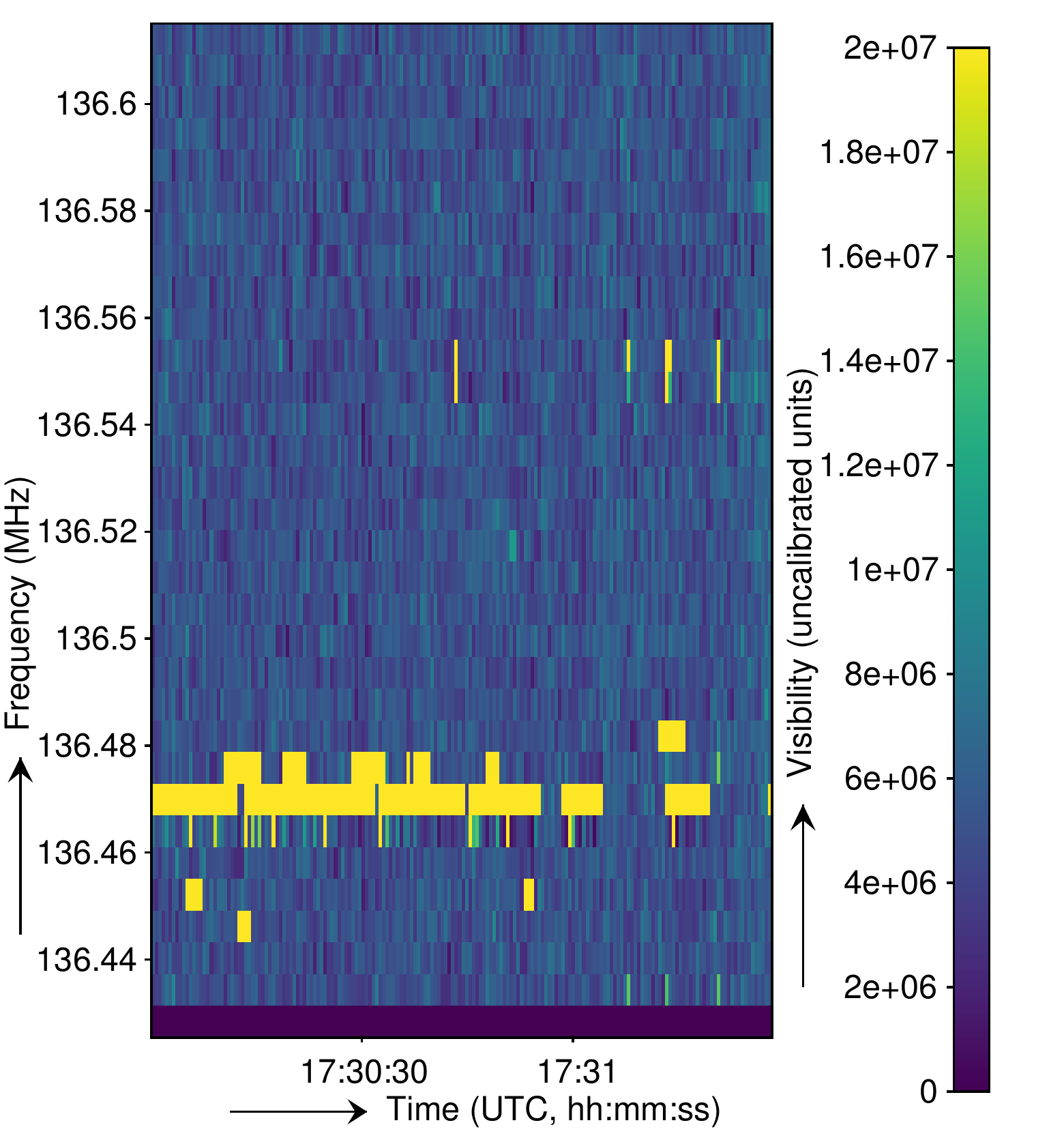}}\hspace*{5mm}%
\subfloat[Row normalization (difference)]{\label{fig:normalization-and-rfi-example-row}\includegraphics[height=7cm]{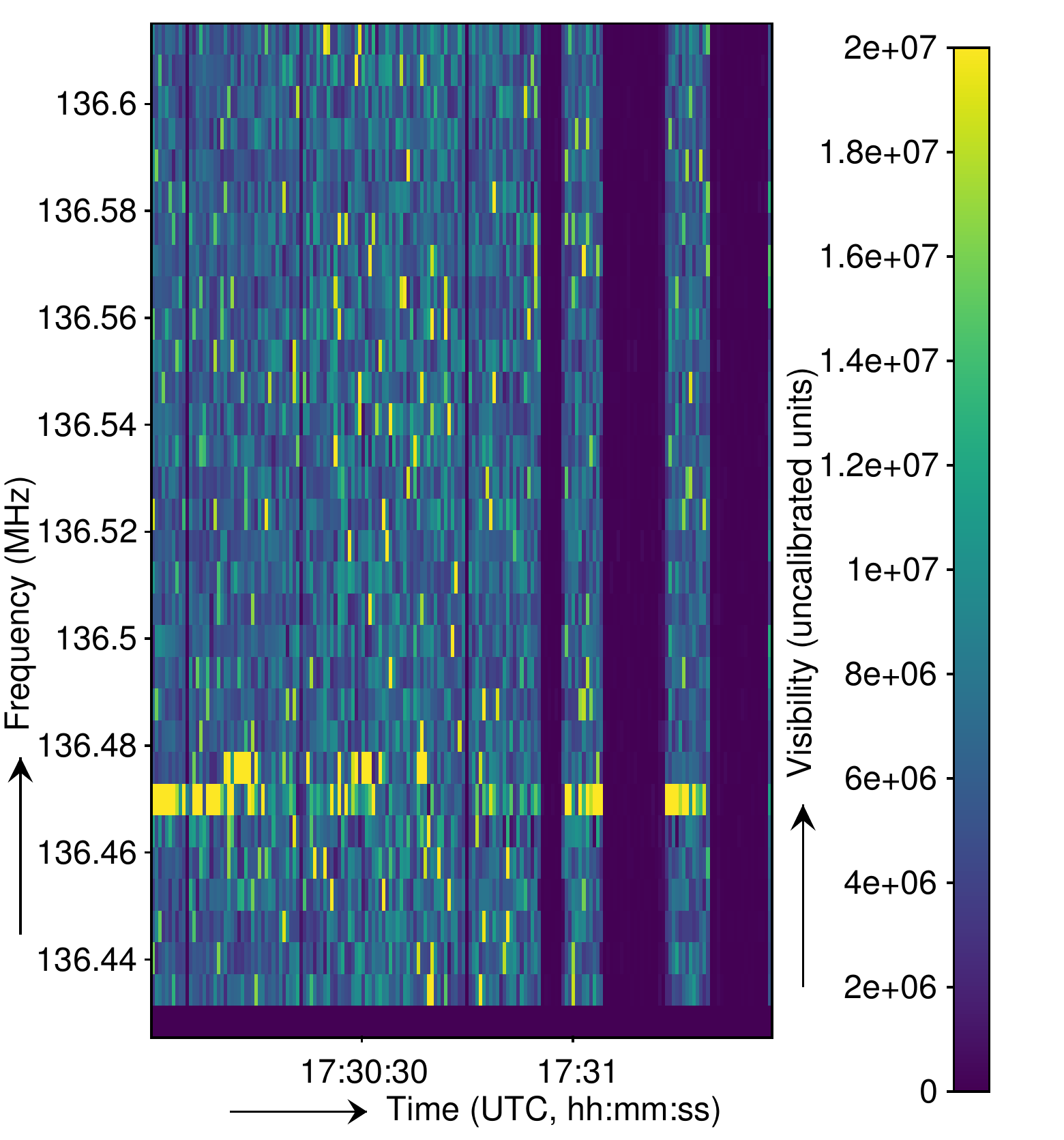}}\\%
\subfloat[AF-normalization (difference)]{\includegraphics[height=7cm]{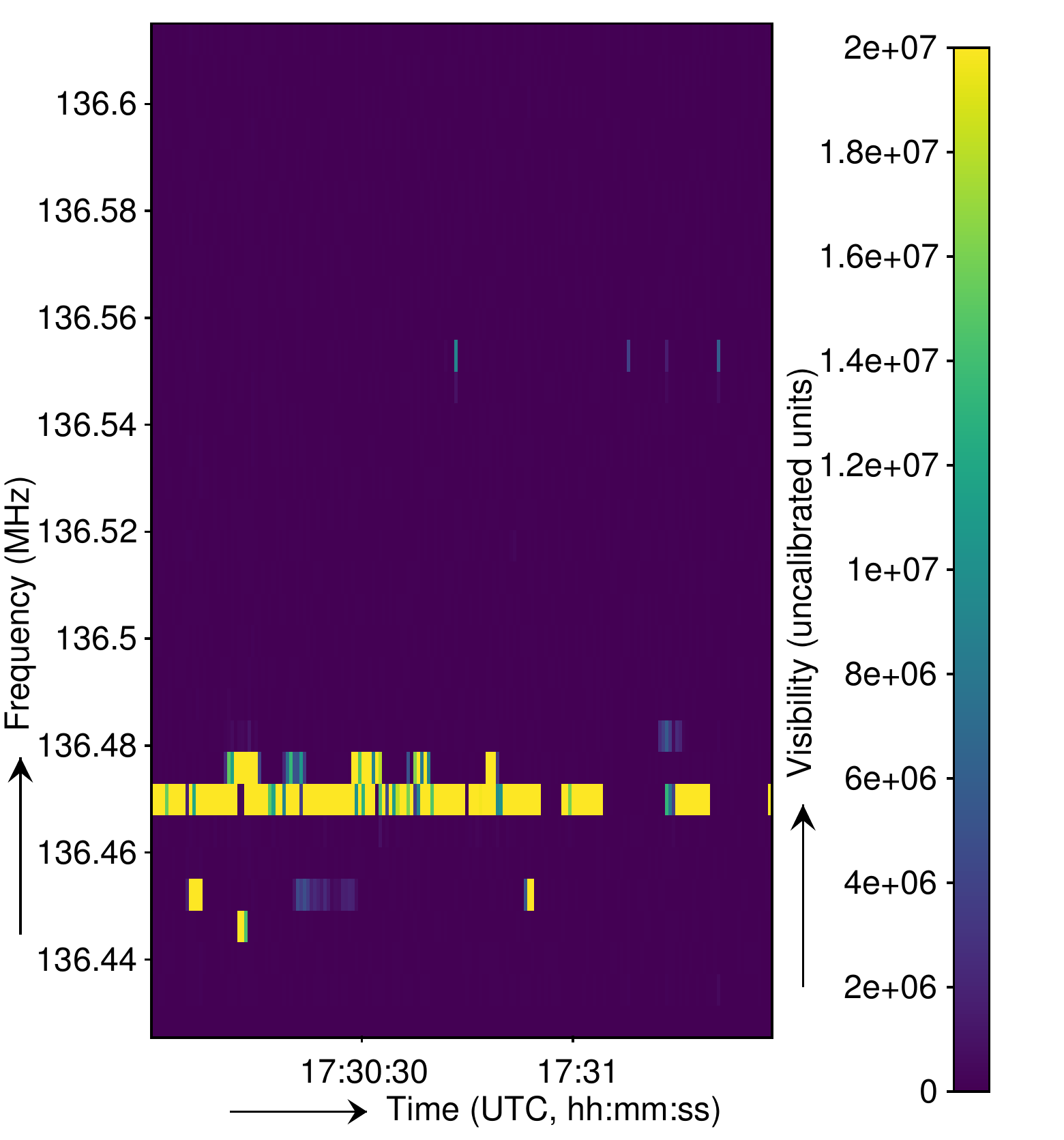}}\hspace*{5mm}%
\subfloat[RF-normalization (difference)]{\includegraphics[height=7cm]{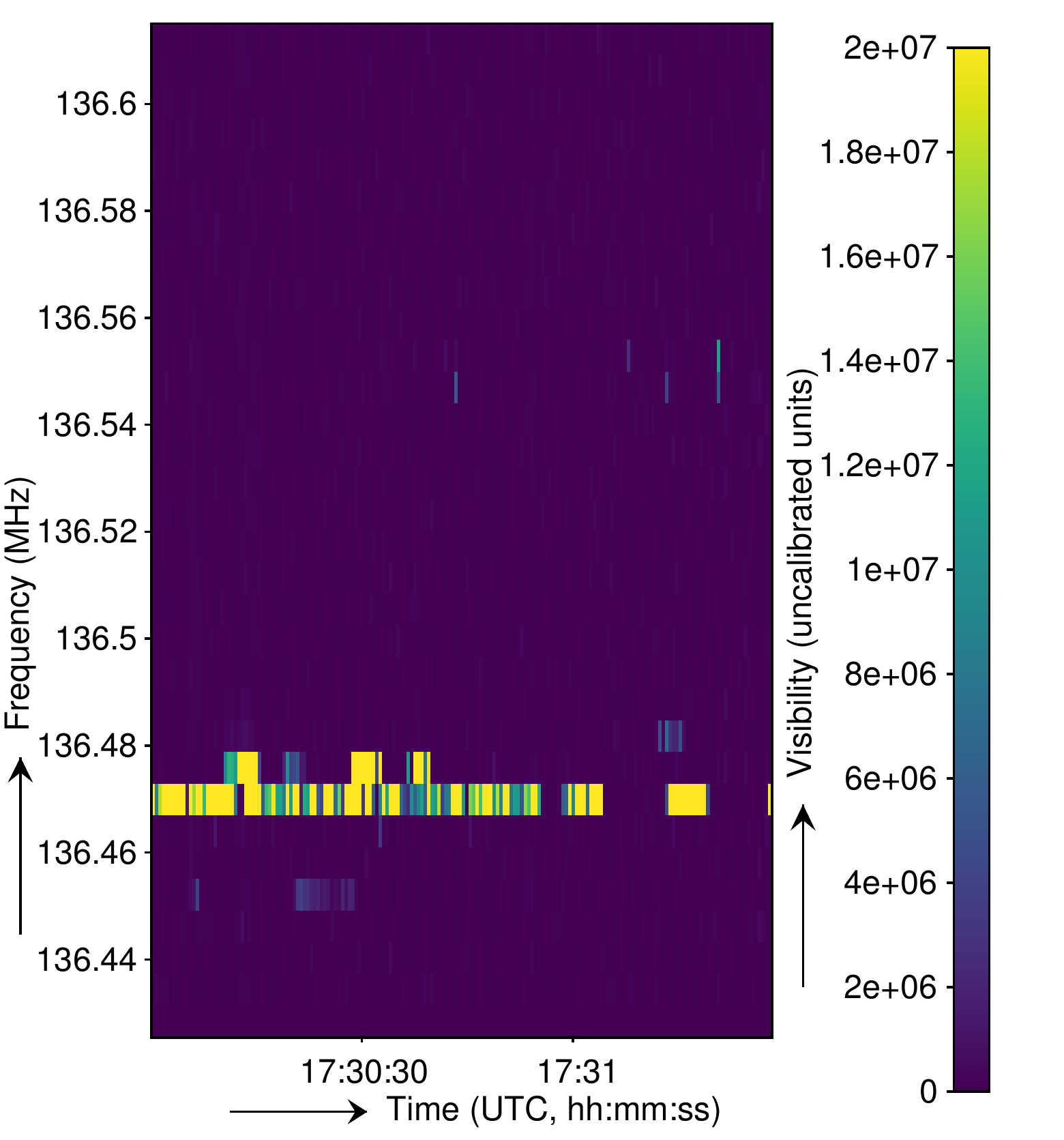}}%
\caption{Dynamic spectra of 1.5 minutes of a LOFAR observation at high frequency and time resolutions (6 kHz and 0.5 s) with strong RFI, compressed to 8 bits per float with different kinds of normalization. Compression with per-row normalization increases the noise significantly in all channels. The RF and AF normalizations, which normalize the channels individually before encoding, behave more robust in the presence of RFI.}
\label{fig:normalization-and-rfi-example}%
\end{figure*}

\begin{figure*}
\centering
\includegraphics[width=7.5cm]{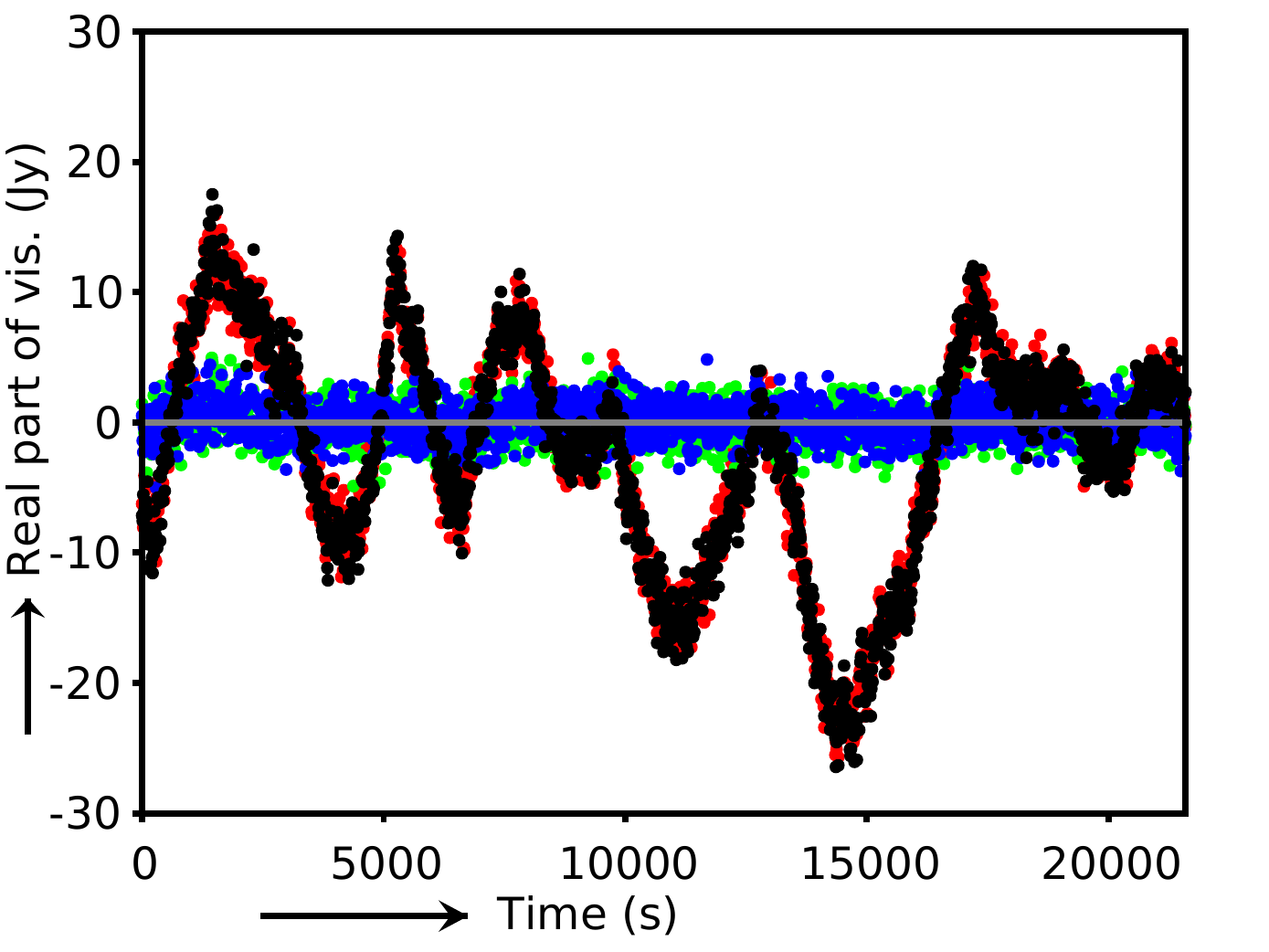}%
\includegraphics[width=7.5cm]{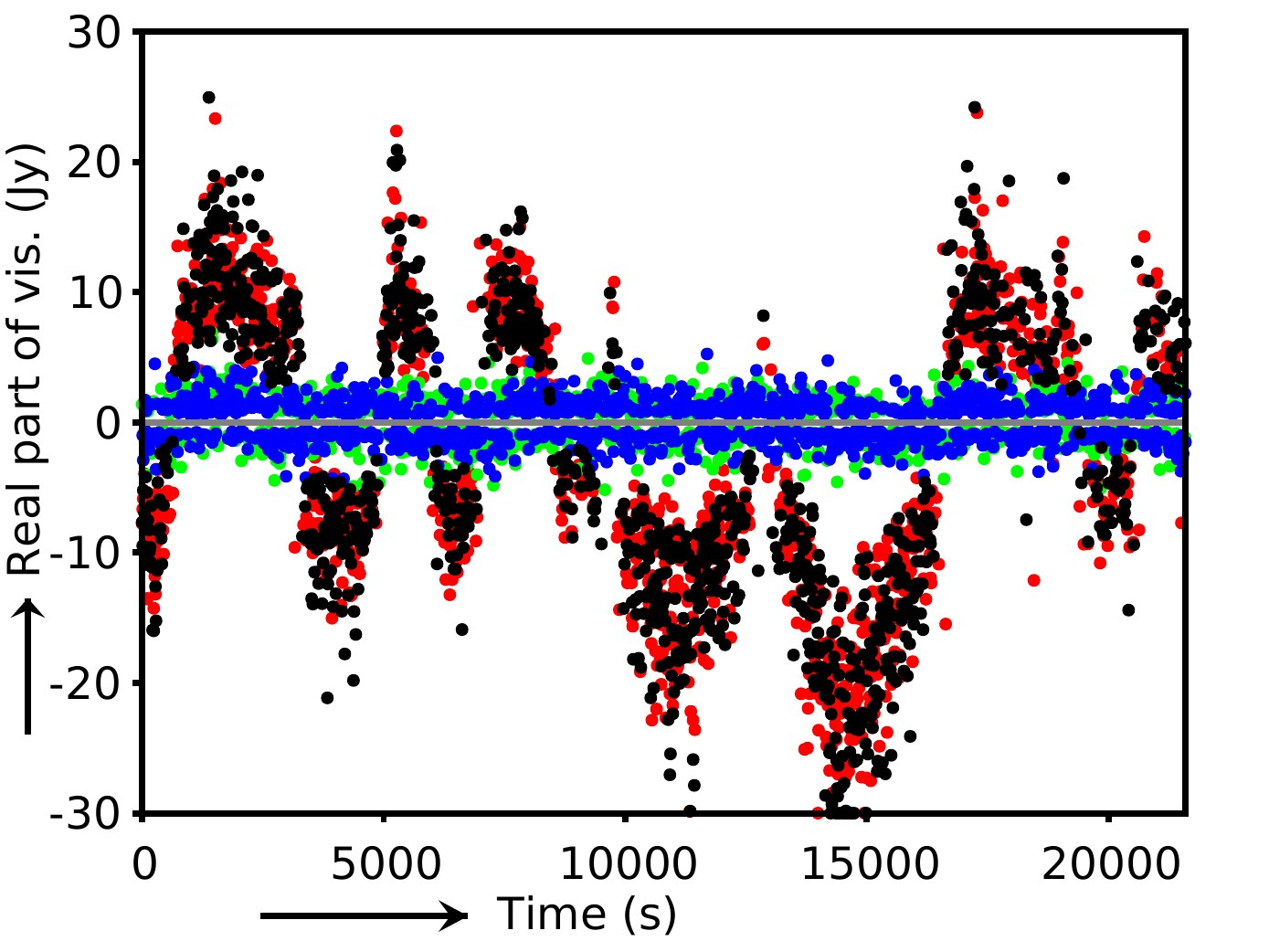}\\%
\hspace*{-0mm}\includegraphics[width=7.5cm]{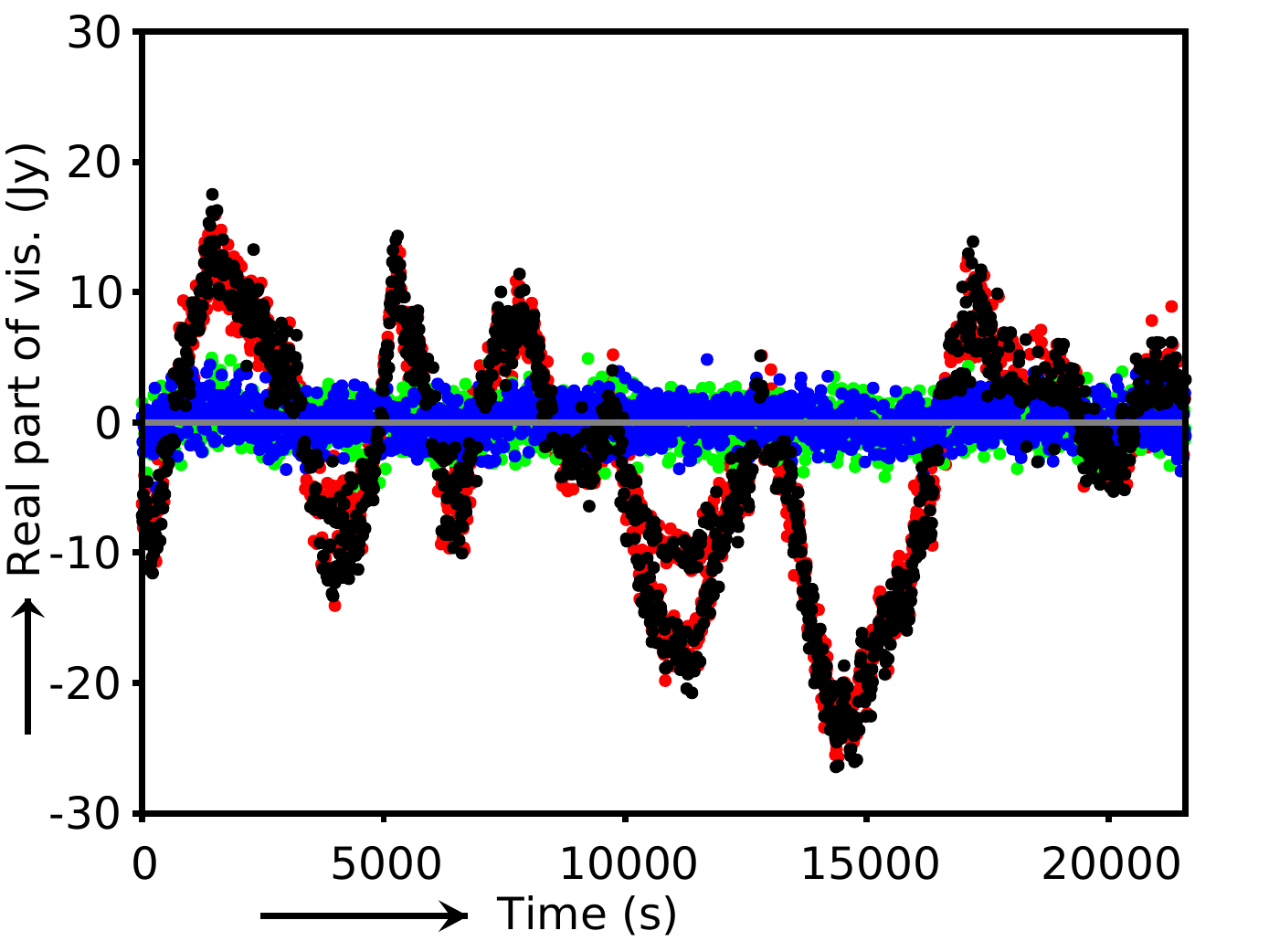}%
\includegraphics[width=7.5cm]{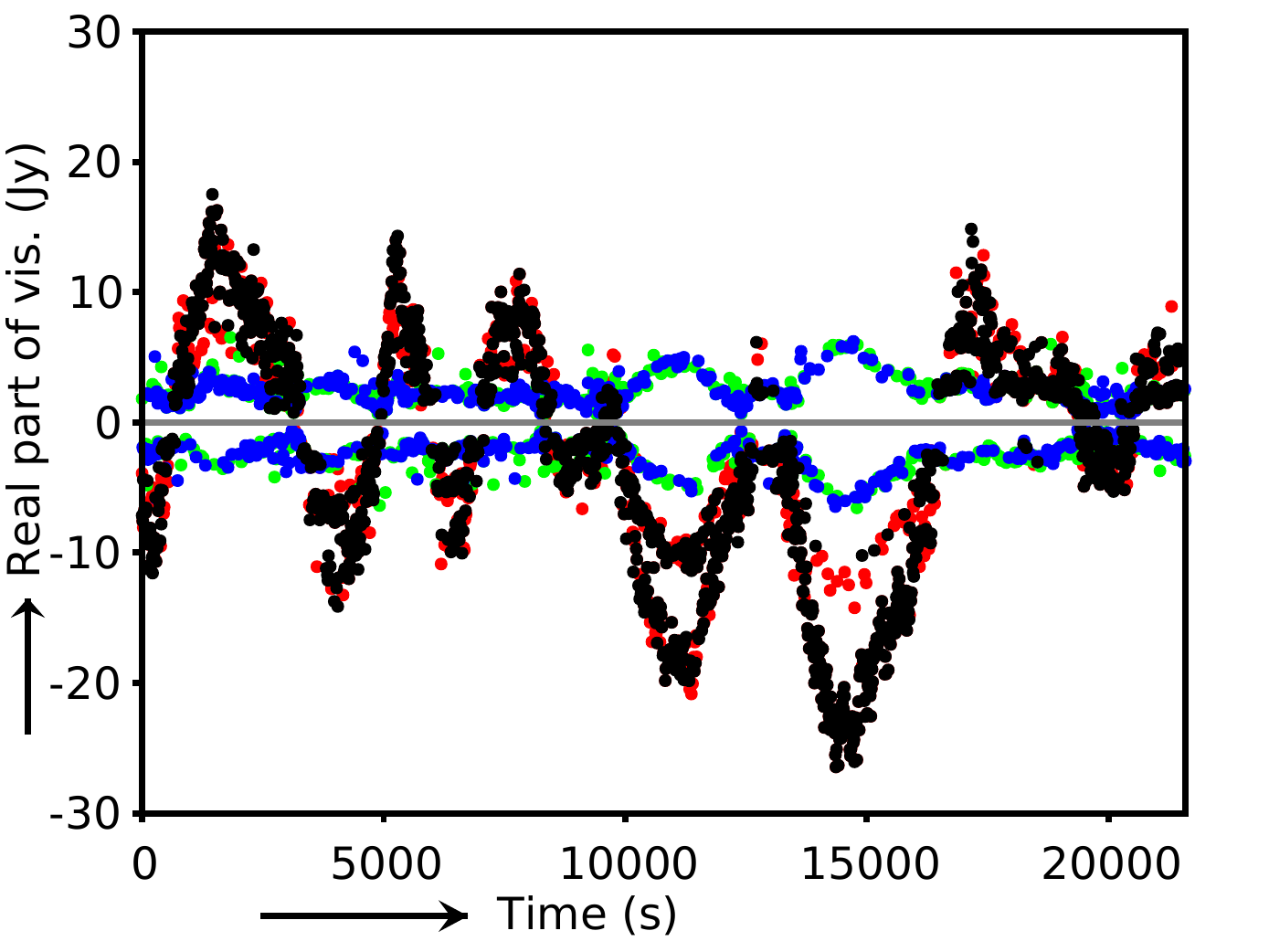}%
\caption{Compression example of a single baseline of a LOFAR observation with high SNR. Top-left plot: original data; top-right plot: compression with AF-normalization; bottom-left plot: compression with RF-normalization; bottom-right plot: compression with row normalization. Compression was performed with the 2.5$\sigma$-truncated Gaussian 3-bit quantization. The red, green, blue and black dots represent the XX, XY, YX and YY correlations, respectively.}
\label{fig:normalization-examples-scatter}%
\end{figure*}

In this work, I describe a new compression method that is particularly designed for noise-dominated visibilities. It makes use of the Gaussian distribution of noise. I will test the effects of this compression method and will determine the maximal compression factor for this technique given a maximal increase in the system noise. In \S\ref{sec:methods}, I will describe the new technique. \S\ref{sec:implementation} presents the implementation. The result of this technique are shown in \S\ref{sec:results} using four test sets. The results are discussed in \S\ref{sec:discussion} and suggestions for future work are given in \S\ref{sec:futurework}. Final conclusions are presented in \S\ref{sec:conclusions}.

\section{Methods} \label{sec:methods}
The new method consists of two steps: a normalization step, in which grouped visibilities are normalized to have a similar distribution; and a quantization and encoding step, which rounds values to a given quantization scheme using a dithering scheme. The new compression technique consisting of the combination of these two steps is named ``Dynamical Statistical Compression'' (Dysco). In the next subsections, we will describe the two steps in detail.

\subsection{Normalization}
In observations, the variance of the data can be different for visibilities of different times, antennae, polarizations and frequencies. To encode visibilities accurately, this variation in variance should be normalized so that the visibilities follow the same distribution. The AIPS uvfits format allows compression of visibilities, and in this format visibilities are normalized by row: each row of visibilities is scaled by a single scaling factor to match the data to the quantization. A row contains an array of visibilities with different frequencies and polarizations, but of the same timestep and antennas \citep{aips-fits-format-2016}. This normalization method will be referred to as ``row normalization''. The downside of this method is that visibilities from different frequencies and polarizations are assumed to follow the same noise distribution. This can be an issue when RFI has not been flagged before encoding, as well as when the bandpass has a large dynamic range or when the statistics of the differently-polarized visibilities differ strongly.

Two additional ways to normalize the visibilities are tested. For the second method, each visibility is normalized by three terms: $f_\textrm{ch}$, a channel scaling term, and $f_\textrm{a}$ and $f_\textrm{b}$, a scaling term for the antennae that were correlated. Hence, for a given visibility, the corresponding normalization factor $\sigma$ can be calculated with $\sigma = f_\textrm{ch} f_\textrm{a} f_\textrm{b}$. Different timesteps and polarizations are encoded independently, so will have different normalization factors, while the real and imaginary values of a visibility are assumed to follow the same distribution. Consequently, for the visibilities corresponding to a given timestep and polarization, $N_\textrm{ch} + N_\textrm{ant}$ factors are stored, as well as $2 N_\textrm{ch} N_\textrm{ant} \left( N_\textrm{ant} + 1 \right) / 2$ quantized values. Such a group of visibilities will be referred to as a timeblock, and this way of normalization as antenna-frequency (AF) normalization. For each timeblock, the channel and antenna factors are stored separately as 32-bit floating point values. 

AF-normalization does not work well on auto-correlations, because these have different statistics, and its standard deviation can generally not be described by the multiplication of two antenna factors. Auto-correlations of astronomical interferometers are seldomly used, and therefore in this work the auto-correlations are set to zero and are ignored during compression. If auto-correlations are to be saved, a separate scaling factor could be computed for auto-correlations. Our implementation does not support this at this point. The other normalization methods do preserve the auto-correlations.

\begin{figure*}[htbp]
\centering
\includegraphics[height=8cm]{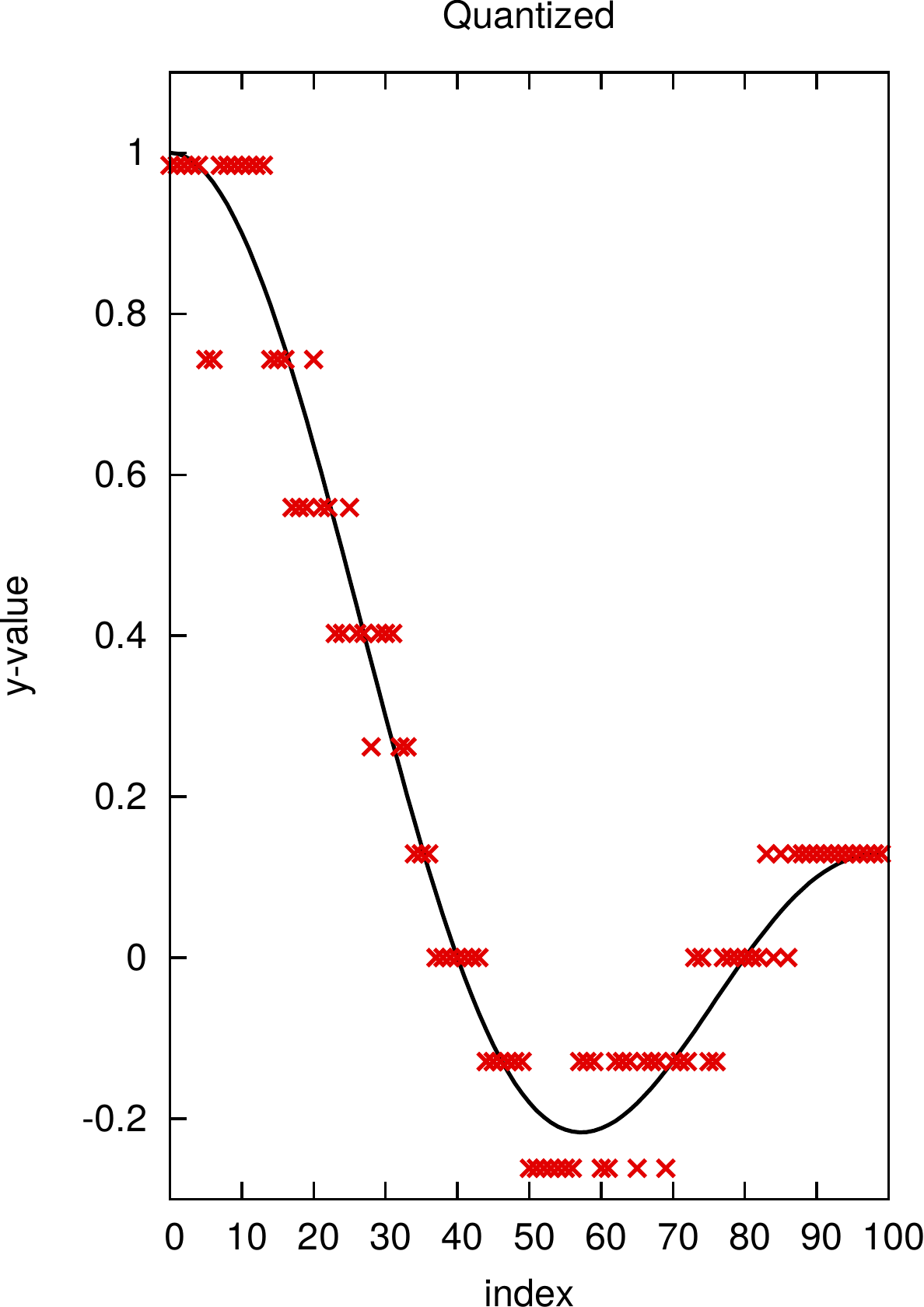}%
\includegraphics[height=8cm]{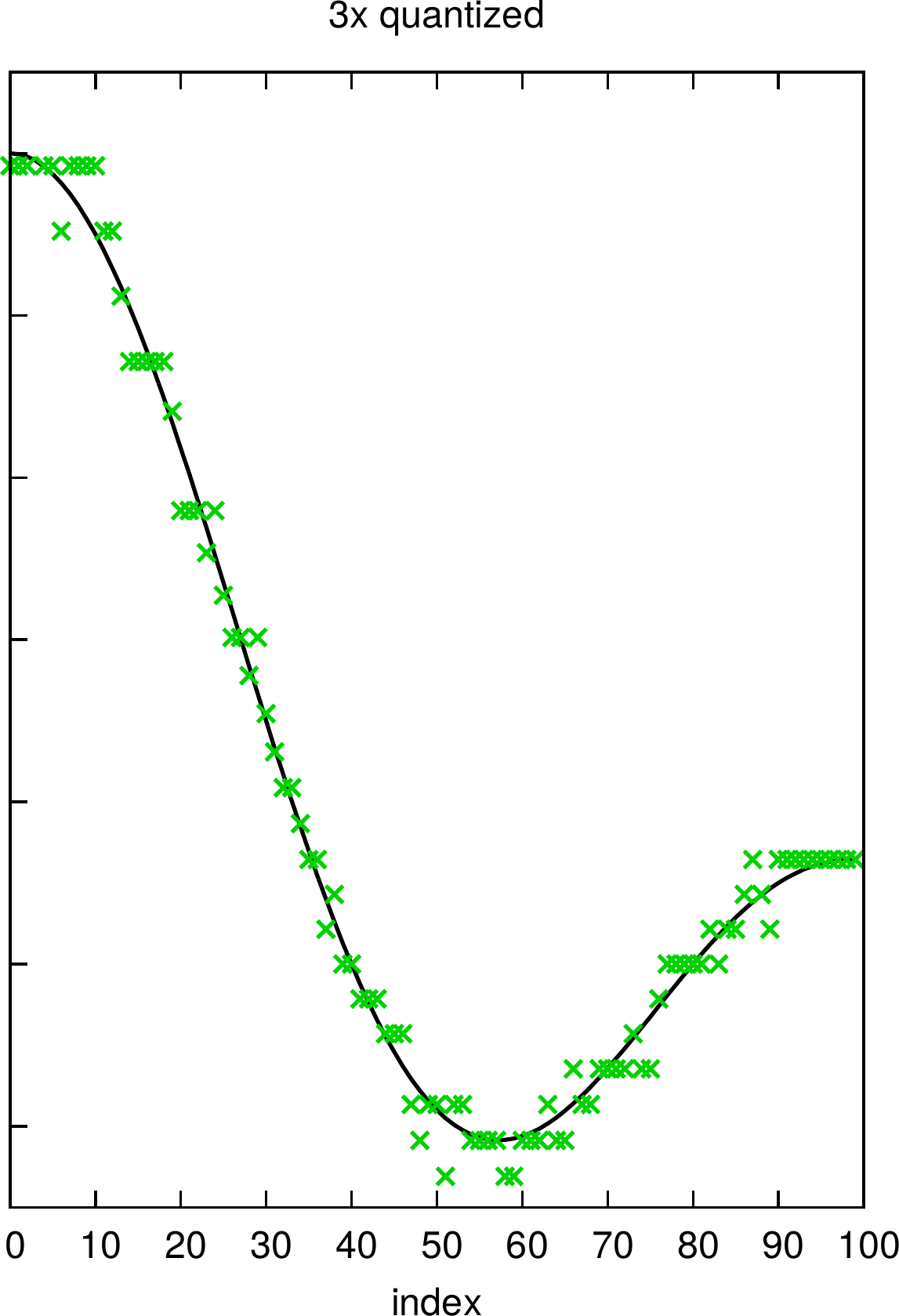}%
\includegraphics[height=8cm]{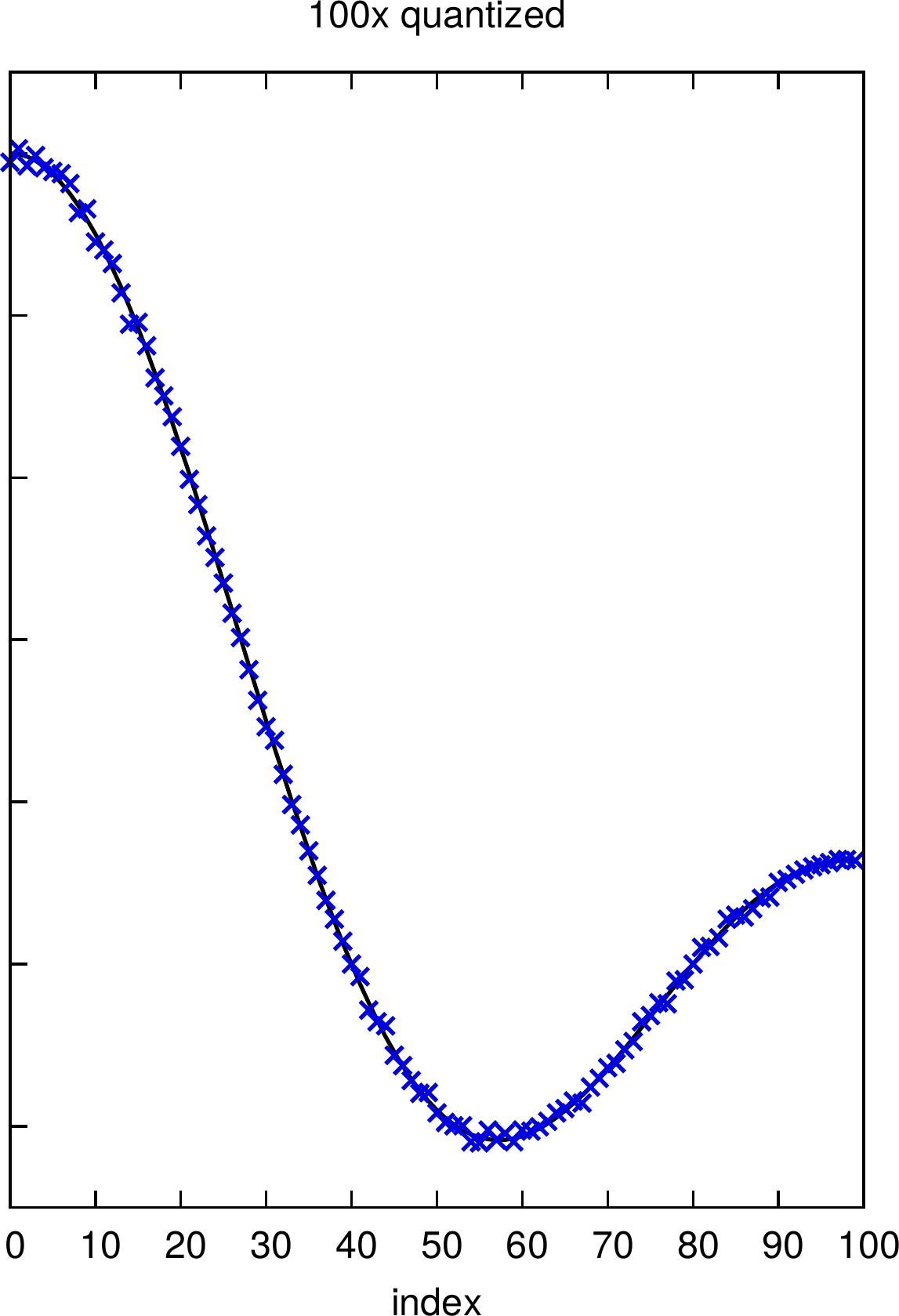}
\caption{A quantization example using the Gaussian-optimized least-squares quantization scheme with dithering to quantize a sinc function. 4-bit quantization and a single scaling factor were used. Left plot: result of encoding and decoding. Because the quantization is optimized for Gaussian distributed values, the quantization steps are smaller near zero. Central  and right plot: average of 3 and 100 times encoding and decoding respectively.}
\label{fig:quantization-example}%
\end{figure*}

The third normalization method uses two terms: $f_\textrm{ch}$, a channel scaling term and $f_\textrm{row}$, a row scaling term. Each polarization also has its own channel and row factors. The final normalization factor is $\sigma = f_\textrm{ch} f_\textrm{row}$. This is somewhat similar to the row normalization method (as used in AIPS) but with some differences: an additional normalization term per frequency is added and independent per-row factors for different polarizations are stored, because the polarizations can have different distributions.

For each of these methods, the data is normalized such that the maximum normalized value is equal to the maximum quantizable value, i.e., the maximum value can be represented exactly and never needs trimming. An alternative is to normalize the data such that a certain range (e.g. all data < 3$\sigma$) fits inside the quantization, and all excess values are trimmed. Initially, for this work the latter method was tested as well. In the case of noise-dominated data, it was found that trimming does not cause a bias, but in case some part of the data has excess values due to real signal, it causes a bias. Hence, trimming data is not considered in this work. Normalization scales the data to fit within the quantization minimum and maximum values. 

In typical situations, in particular for data outputted by the LOFAR or MWA correlators, the extra metadata (the scaling factors) stored by any of the normalization methods, is negligible.

The result of these normalization methods is demonstrated in Fig.~\ref{fig:normalization-and-rfi-example} for an observation with strong RFI. The original spectrum is shown in the top-left. The other dynamic spectra show the difference between before and after compression using the three normalization techniques. With per-row normalization, when one of the channels contain large values, the noise in the other channels increases strongly. With AF and RF normalization, which have an extra channel normalization factor, the noise in the data without RFI is not significantly increased.

Fig.~\ref{fig:normalization-examples-scatter} shows another example of the properties of normalization. These scatter plots show a single baseline of a LOFAR observation. This particular observation was averaged to a resolution of 12 second and 200 kHz, and because of this it has a high SNR in the XX and YY polarizations. All three normalizations visibly add noise to the data, but in different ways. The assumption for AF-normalization is that the data is noise dominated. Because this is not the case in this example, the noise added to the XX and YY polarizations is larger than for the other normalizations. Row normalization does not store an independent scaling factor per polarization. Because of this, the accuracy of the XY and YX quantization strongly depends on the XX and YY signals. It is important to note that the quantization is unbiased; while the quantization adds certain structures to the raw visibilities (most clearly visible in row normalization), it will be shown in later sections that the added imaging noise is unstructured and unbiased.

\subsection{Quantization}
\begin{table*}[htbp]
\caption{Sets used for testing the compression.}\label{tbl:obs-list}
\begin{tabular}{|l|l|c|c|c|c|l|}
\hline
\textbf{Set} & \textbf{Observation} & \textbf{id} & \textbf{date \& size} & \textbf{$\Delta t$} & \textbf{$\Delta \nu$} & \textbf{Description, processing} \\
\hline
A   & LOFAR HBA \object{3C 196}, & L431602 & 2016-02-28 & 4 s & 36 kHz & Single subband, calibrated after\\
    & 6 hour, 117 MHz  & & 6.1 GB     & & (5 ch) & compression. \\
\hline
B   & LOFAR HBA 3C 196, & L323164 & 2015-04-01 & 0.5 s & 6 kHz & Single subband, calibrated before \\
    & 10 min, 133 MHz  & & 6.0 GB     & & (32 ch) &  compression. \\
\hline
C   & MWA \object{Vela} / \object{Puppis A}, & 1052736496 & 2013-05-16 & 4 s & 80 kHz  & Calibrated before compression. \\
    & 2 min, 149 MHz   & & 2.3 GB     & & (384 ch)& \\
\hline
D   & MWA \object{Hydra A},        & 1077974936 & 2014-03-04 & 0.5s& 40 kHz  & Calibrated after compression, recompressed \\
    & 2 min, 154 MHz   & & 65 GB      & & (768 ch)& after applying solutions. \\
\hline
\end{tabular}
\end{table*}

After normalization, the visibilities are encoded with a non-linear quantization scheme. The encoding is a-priory optimized for complex samples with a given distribution with zero mean. The distribution of correlated samples will approximate a normal distribution when the SNR is low and the observation does not contain interfering sources. Several encoding schemes are tested that are optimized for different distributions. 

The quantization table is created by uniform sampling of the inverse of the corresponding cumulative distribution function. Dithering is applied to avoid possible systematic bias when the signal-to-noise ratio in the input data is high. Dithering is implemented as follows: For each float to be encoded, the two closest quantization values are found. The chosen encoding quantity is selected randomly, giving each a probability dependent on its distance to the original value. A quantized value will always be selected when it equals the original value, while if the original value is halfway between the two quantized values, both quantities are equal likely to be selected. This is equivalent to adding uniformly-distributed random noise to the input value, which is a common method of applying dithering, except that the bounds of the distribution vary per sample, depending on the distance of the value to the quantization values. Dithering ensures that the average of encoding a value many times asymptotically approaches that value. An example of the quantization is shown in Fig.~\ref{fig:quantization-example}.

The quantized values are converted to binary values and bit-packed. The number of bits can be set by the user, and define the number of quantities per encoding symbol. One encoding quantity is reserved for storing the special value `not-a-number' (NaN). An odd number of quantities remains for encoding finite values. Since the quantized values are symmetric around zero, the value 0 can be encoded precisely. This is beneficial, because it implies that no noise will be added when encoding the value zero. This way, an antenna that has produced zeros can be more easily identified. As an example, when the user asks for 8 bit encoding, a visibility is quantized to 256 values: values <127 correspond with negative visibilities; 127 corresponds with zero; values >127 and <255 correspond with positive visibilities; and 255 corresponds with the value NaN.

\section{Implementation} \label{sec:implementation}
The Casacore table system that is used to access CASA measurement sets uses so-called `storage managers' to access visibilities on disk. The storage manager system is transparent to the client program, and storage managers can be located in an external software library. Any client program that uses Casacore (including CASA) can access measurement sets that were stored with a custom storage manager, without any changes to or recompilation of the client software. For this project, such a Casacore storage manager was designed to allow the transparent compression of data in a Casacore measurement set with the Dysco technique. The source code of the Dysco implementation is publicly released\footnote{The source code and manual of the implementation can be found at \url{https://github.com/aroffringa/dysco}. Commit 393cd9775875f779d0d0cc8944d6868c05a98345 (8 Aug 2016) was used.}.

In my implementation, the compressed data is stored in an extra file inside the measurement set. Inside the file, the timeblocks are stored consecutively. Given that a timeblock has a fixed number of visibilities, as well as that the encoded representation has a fixed size, the position of a timeblock inside the file can be directly calculated. This has implementational advantages, such as not having to maintain an index table, as well as being able to replace a timeblock without the possibility that the new timeblock will not fit.

Because each timeblock is compressed independently in each of the normalization and quantization methods, random access and streaming access of the timeblocks is possible. When a timeblock is requested or written, the implementation will keep that timeblock in the cache until another timeblock is written or read. This avoids encoding or decoding the timeblock when the client accesses different visibilities (table rows) from the same block multiple times. Once a different timeblock is requested, the cached timeblock is queued for writing if it has been changed. This system implies that accessing the visibilities in time-major order requires encoding or decoding each time block only once. This access pattern is the most common way of accessing a measurement set --- calibration and imaging access the data in this way\footnote{One exception is flagging \citep{lofar-radio-environment}, for which a reordering of the data is performed.}. Random access to measurement sets will lead to encoding or decoding the same block multiple times, but in practice this access pattern is rare.

AF-normalization is implemented as follows: first, the channel standard deviations are divided out to make each channel have a standard deviation of unity. Then, the antenna factors are found by solving $\Sigma = \mathbf{f}_\textrm{ant} \otimes \mathbf{f}_\textrm{ant}$, where $\otimes$ is the outer product, $\Sigma$ is a matrix with the measured variance of a baseline at each element after channel normalization, and $\mathbf{f}_\textrm{ant}$ is a vector with the antenna factors to be solved for. After dividing out the antenna factors, such that the variance of each baseline is approximately unity, the algorithm continues by iteratively selecting and maximizing the channel factor or antenna factor that increases the sum of the absolute values the most when maximized. This is continued until the gain is smaller than some threshold.

In the RF-normalization algorithm, the standard deviations of the channels are also first normalized, but afterwards each row is directly maximized. After that, each channel is maximized.

The encoding of timeblocks is multi-threaded. Typical nodes can encode and write visibilities faster than writing the uncompressed visibilities.

\section{Results} \label{sec:results}
\begin{table} \caption{Imaging error RMS values with 8-bit compression. Averages have been calculated by normalizing the results for each set.}\label{tbl:8bitresults}\begin{center}\begin{tabular}{llccc} %
\hline
Set & Quantization & AF & RF & row \\
    &  & RMS & RMS & RMS \\
\hline
A & Gaussian      & 179 $\mu$Jy & 462 $\mu$Jy & 612 $\mu$Jy\\
  & Tr. Gaus. 1.5 & 227 $\mu$Jy &  92 $\mu$Jy & 105 $\mu$Jy\\
  & Tr. Gaus. 2.5 & 178 $\mu$Jy & 209 $\mu$Jy & 265 $\mu$Jy\\
  & Tr. Gaus. 3.5 & 178 $\mu$Jy & 437 $\mu$Jy & 572 $\mu$Jy \\
  & Uniform       & 299 $\mu$Jy &  78 $\mu$Jy &  82 $\mu$Jy\\
\hline
B & Gaussian      & 117 $\mu$Jy & 174 $\mu$Jy & 224 $\mu$Jy \\
  & Tr. Gaus. 1.5 &  98 $\mu$Jy & 212 $\mu$Jy &  75 $\mu$Jy \\
  & Tr. Gaus. 2.5 &  91 $\mu$Jy & 160 $\mu$Jy & 113 $\mu$Jy \\
  & Tr. Gaus. 3.5 & 115 $\mu$Jy & 174 $\mu$Jy & 196 $\mu$Jy \\
  & Uniform       & 128 $\mu$Jy & 290 $\mu$Jy &  82 $\mu$Jy \\
\hline
C & Gaussian      &  421 $\mu$Jy & 452 $\mu$Jy & 406 $\mu$Jy\\
  & Tr. Gaus. 1.5 &  616 $\mu$Jy & 327 $\mu$Jy & 360 $\mu$Jy\\
  & Tr. Gaus. 2.5 &  453 $\mu$Jy & 318 $\mu$Jy & 309 $\mu$Jy \\
  & Tr. Gaus. 3.5 &  422 $\mu$Jy & 459 $\mu$Jy & 394 $\mu$Jy \\
  & Uniform       &  827 $\mu$Jy & 411 $\mu$Jy & 462 $\mu$Jy \\
\hline
D & Gaussian      & 81 $\mu$Jy & 104 $\mu$Jy & 93 $\mu$Jy\\
  & Tr. Gaus. 1.5 & 76 $\mu$Jy &  67 $\mu$Jy & 74 $\mu$Jy\\
  & Tr. Gaus. 2.5 & 68 $\mu$Jy &  70 $\mu$Jy & 70 $\mu$Jy \\
  & Tr. Gaus. 3.5 & 79 $\mu$Jy &  99 $\mu$Jy & 90 $\mu$Jy \\
  & Uniform       & 95 $\mu$Jy &  81 $\mu$Jy & 91 $\mu$Jy \\
\hline
Avg
  & Gaussian      & 199 $\mu$Jy & 305 $\mu$Jy & 344 $\mu$Jy\\
  & Tr. Gaus. 1.5 & 225 $\mu$Jy & 195 $\mu$Jy & 153 $\mu$Jy\\
  & Tr. Gaus. 2.5 & 184 $\mu$Jy & 201 $\mu$Jy & 194 $\mu$Jy \\
  & Tr. Gaus. 3.5 & 197 $\mu$Jy & 296 $\mu$Jy & 320 $\mu$Jy \\
  & Uniform       & 294 $\mu$Jy & 243 $\mu$Jy & 177 $\mu$Jy \\
\hline
\multicolumn{2}{l}{Per-normalization avg} & 220 $\pm$   & 248 $\pm$   & 237 $\pm$ \\
  &                                       &  31 $\mu$Jy &  23 $\mu$Jy &  39 $\mu$Jy \\
\hline
  &  & \multicolumn{2}{c}{Per-quantization avg} \\
  & Gaussian      & \multicolumn{2}{c}{$282 \pm 62$ $\mu$Jy} \\
  & Tr. Gaus. 1.5 & \multicolumn{2}{c}{$191 \pm 46$ $\mu$Jy}\\
  & Tr. Gaus. 2.5 & \multicolumn{2}{c}{$193 \pm 20$ $\mu$Jy}\\
  & Tr. Gaus. 3.5 & \multicolumn{2}{c}{$271 \pm 55$ $\mu$Jy}\\
  & Uniform       & \multicolumn{2}{c}{$238 \pm 69$ $\mu$Jy}\\
\hline\end{tabular}
\end{center}
\end{table}

In this section, Dysco is tested using several real observations. Table~\ref{tbl:obs-list} lists the sets that have been used for testing. LOFAR HBA and MWA observations are used with different time and frequency resolutions. Because the compression method adds noise that is relative to the variance of the data, sets with different signal-to-noise ratios are used for testing. For both telescopes, a high-resolution set was used, which is the typically resolution for archiving, and a set at lower resolution was used that is the typical resolution used during processing. 

Images are produced for each of the sets and for various compression configurations. Reference images are made by skipping the compression step. Typical imaging parameters for the field of interest are used. Two quantities are measured: the RMS of the difference between the reference and the dirty images of the compressed set; and the absolute RMS levels of the images after cleaning. The differential RMS is measured over the entire image. The absolute RMS is measured in a rectangular part of the image in which no sources are visible. Each test set is tested with a different processing strategy, in order to test the effect of compression in combination with different processing steps.

\begin{figure*}
\centering
\hspace*{-6mm}\includegraphics[height=7.5cm]{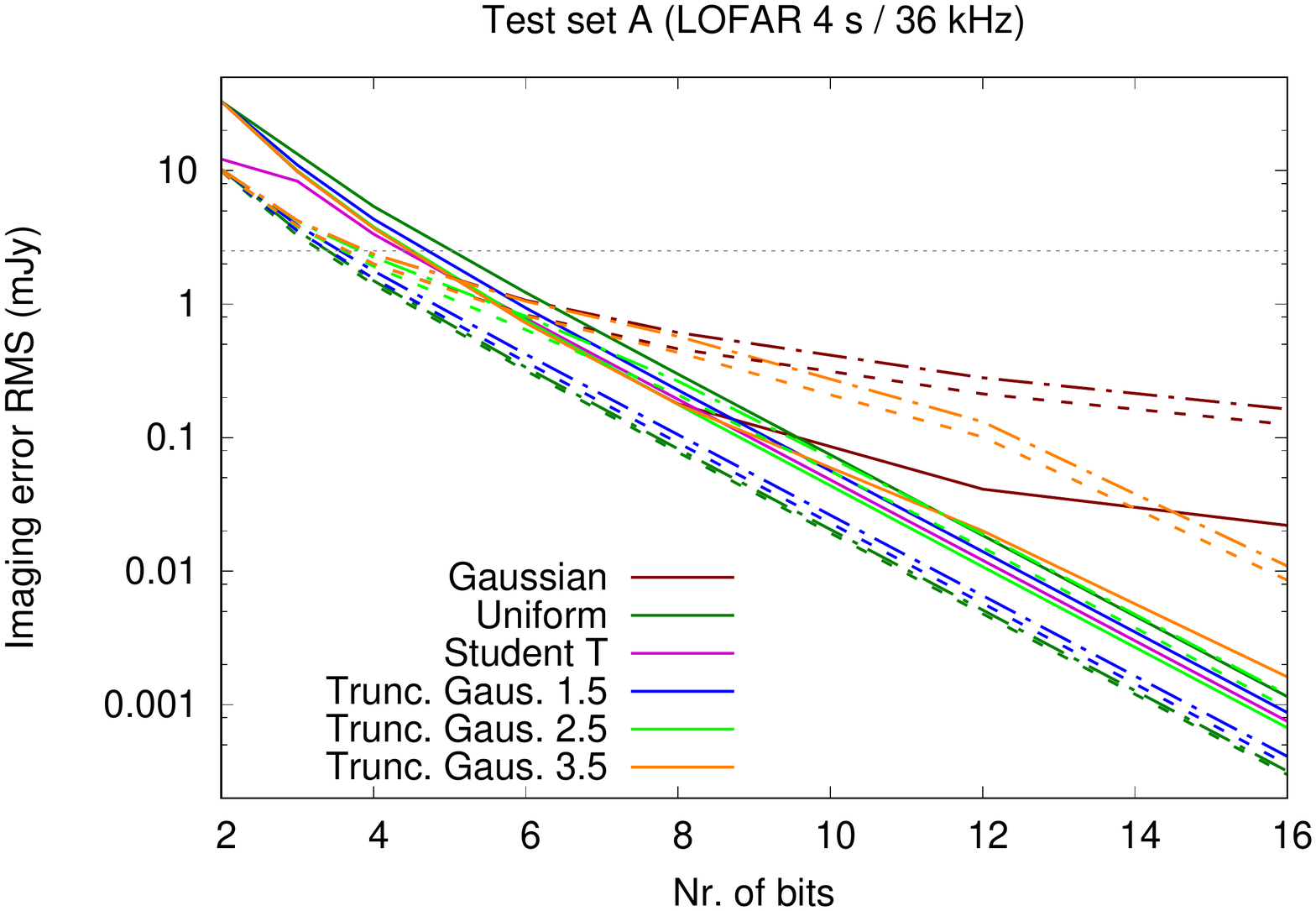}%
\hspace*{-16mm}\includegraphics[height=7.5cm]{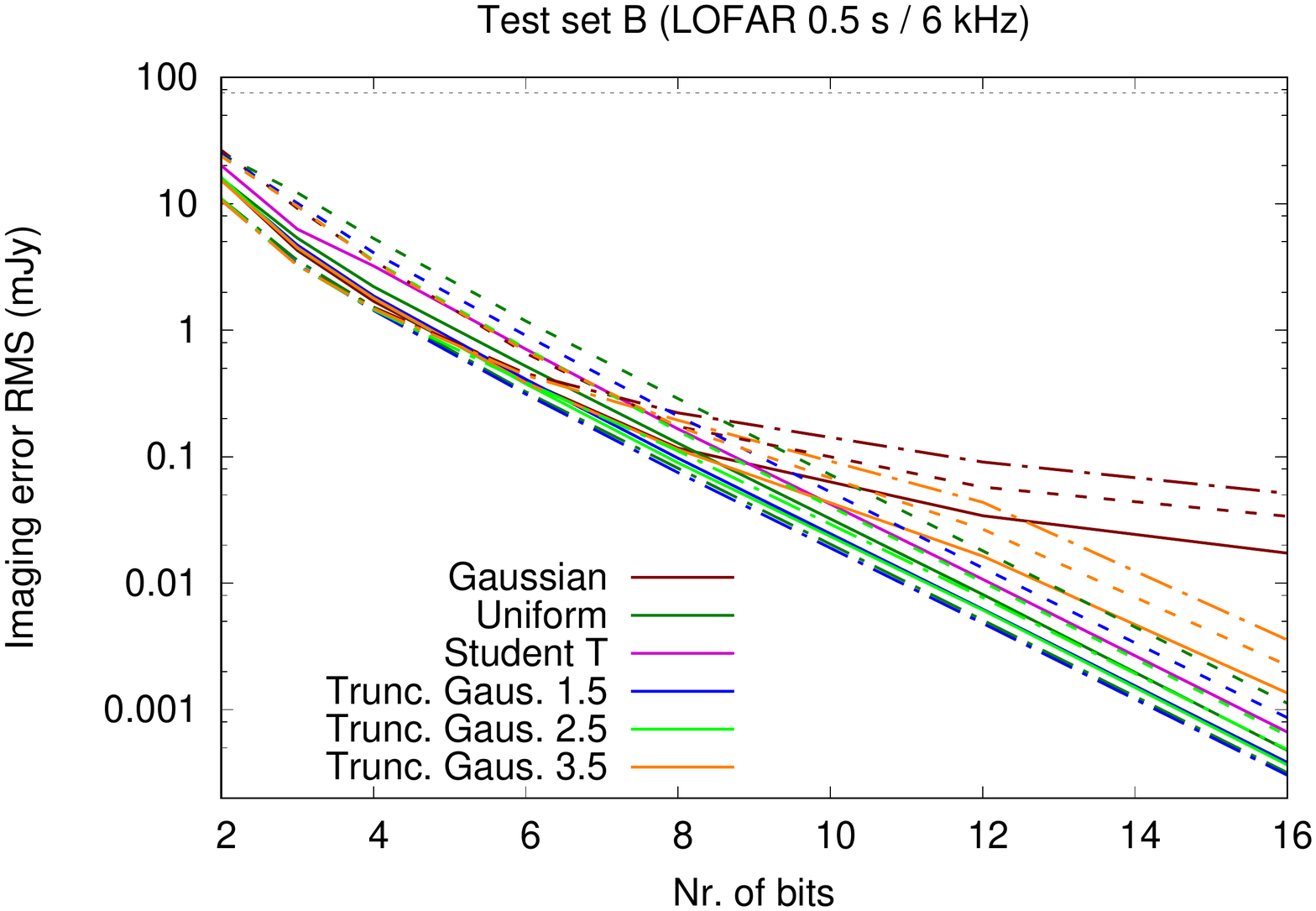}\\%
\hspace*{-6mm}\includegraphics[height=7.5cm]{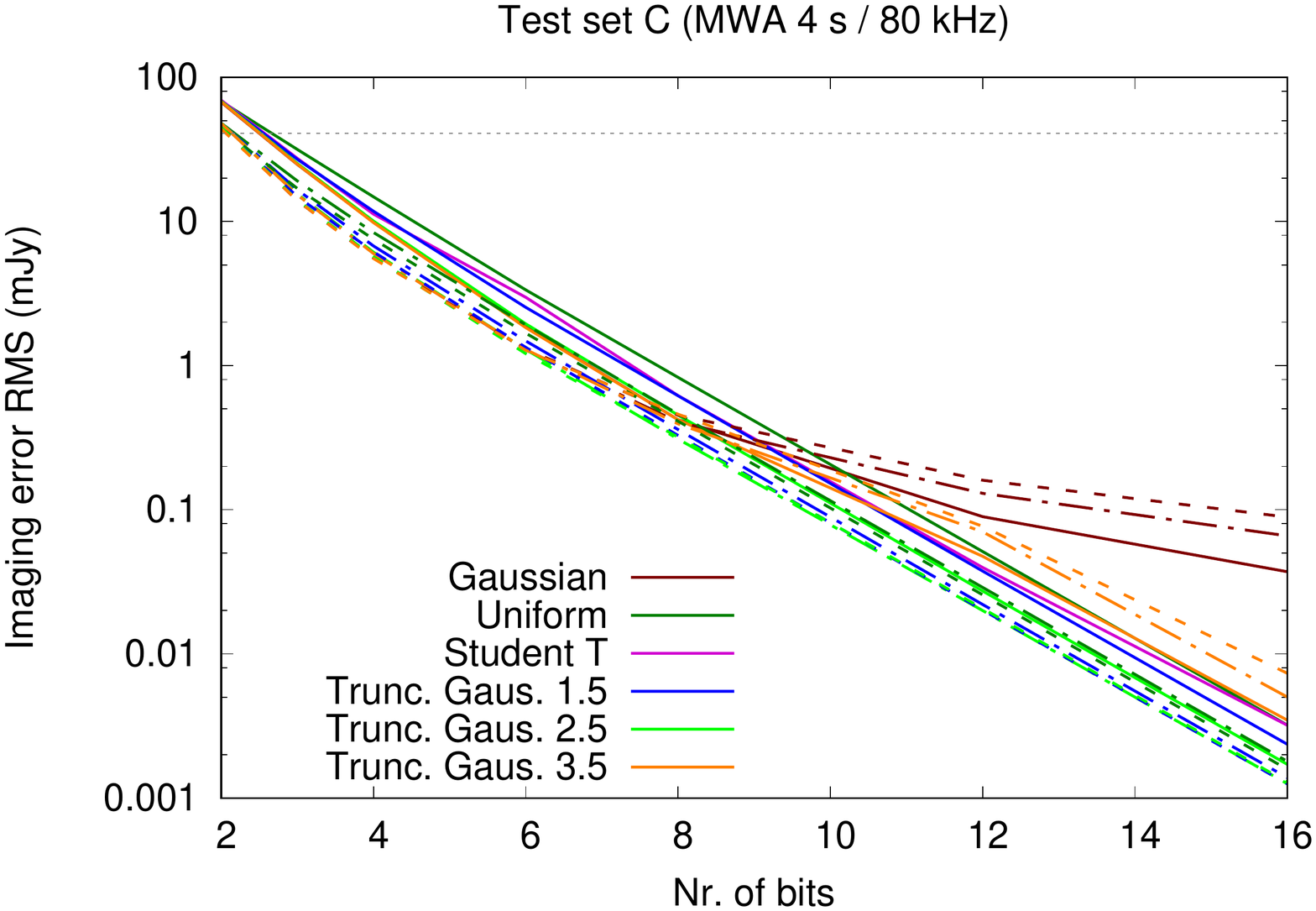}%
\hspace*{-16mm}\includegraphics[height=7.5cm]{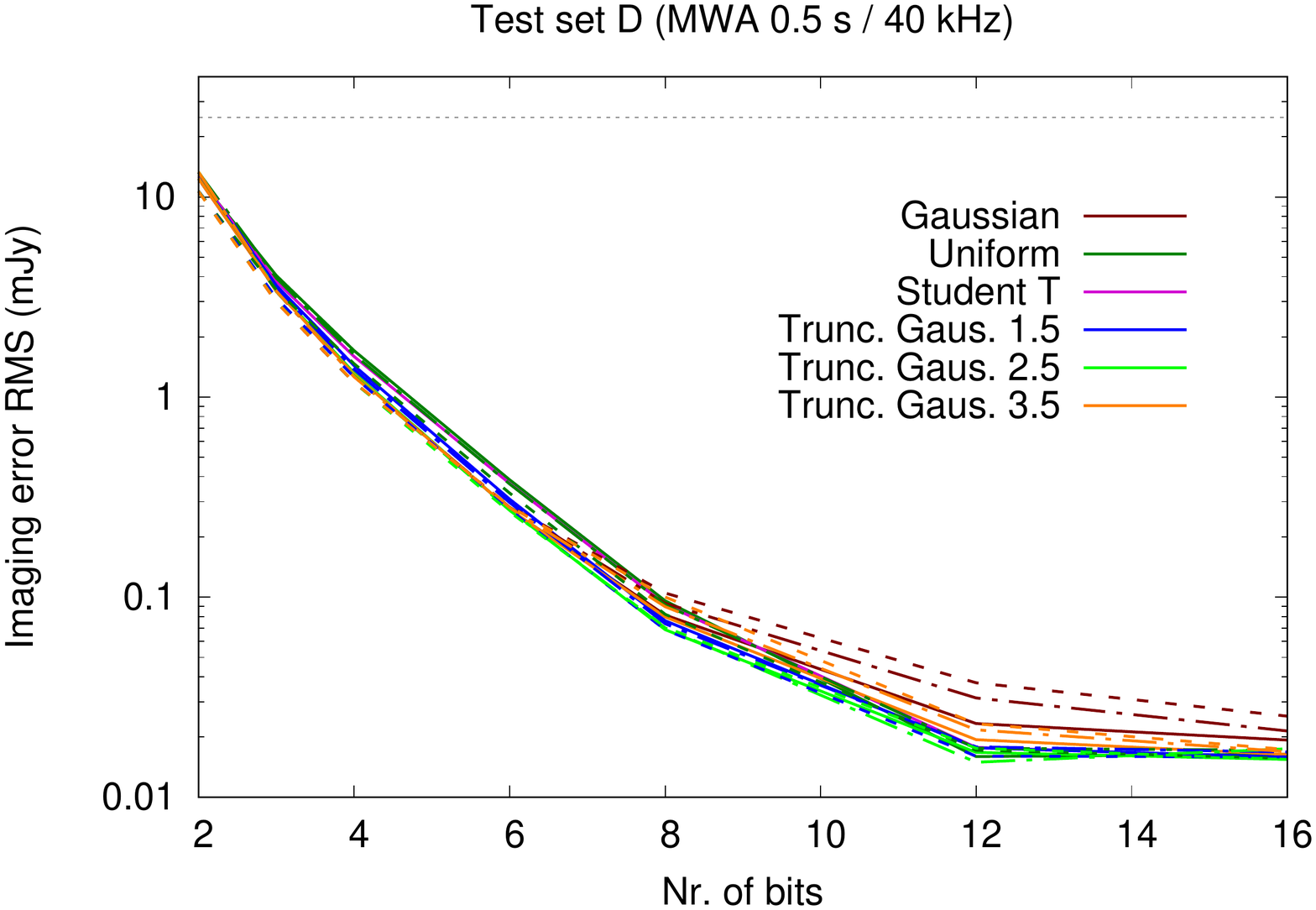}\\%
\caption{Compression noise RMS values for different quantization schemes and test sets, determined by measuring the RMS in the difference between images processed with and without compression. The horizontal dashed grey line indicates the system noise level. }
\label{fig:result-diffnoise}%
\end{figure*}

\begin{figure*}
\centering
\hspace*{-6mm}\includegraphics[height=7.5cm]{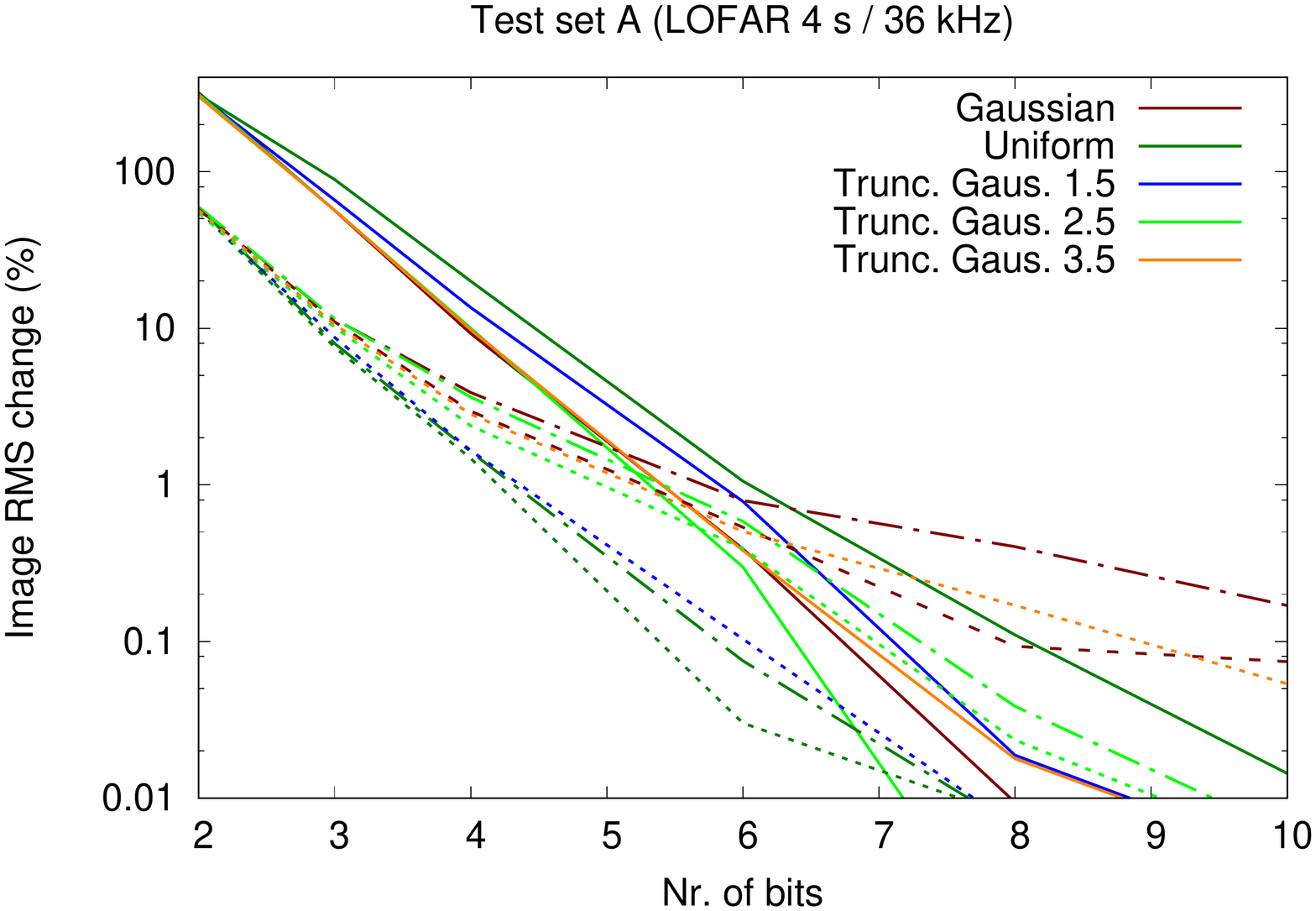}%
\hspace*{-16mm}\includegraphics[height=7.5cm]{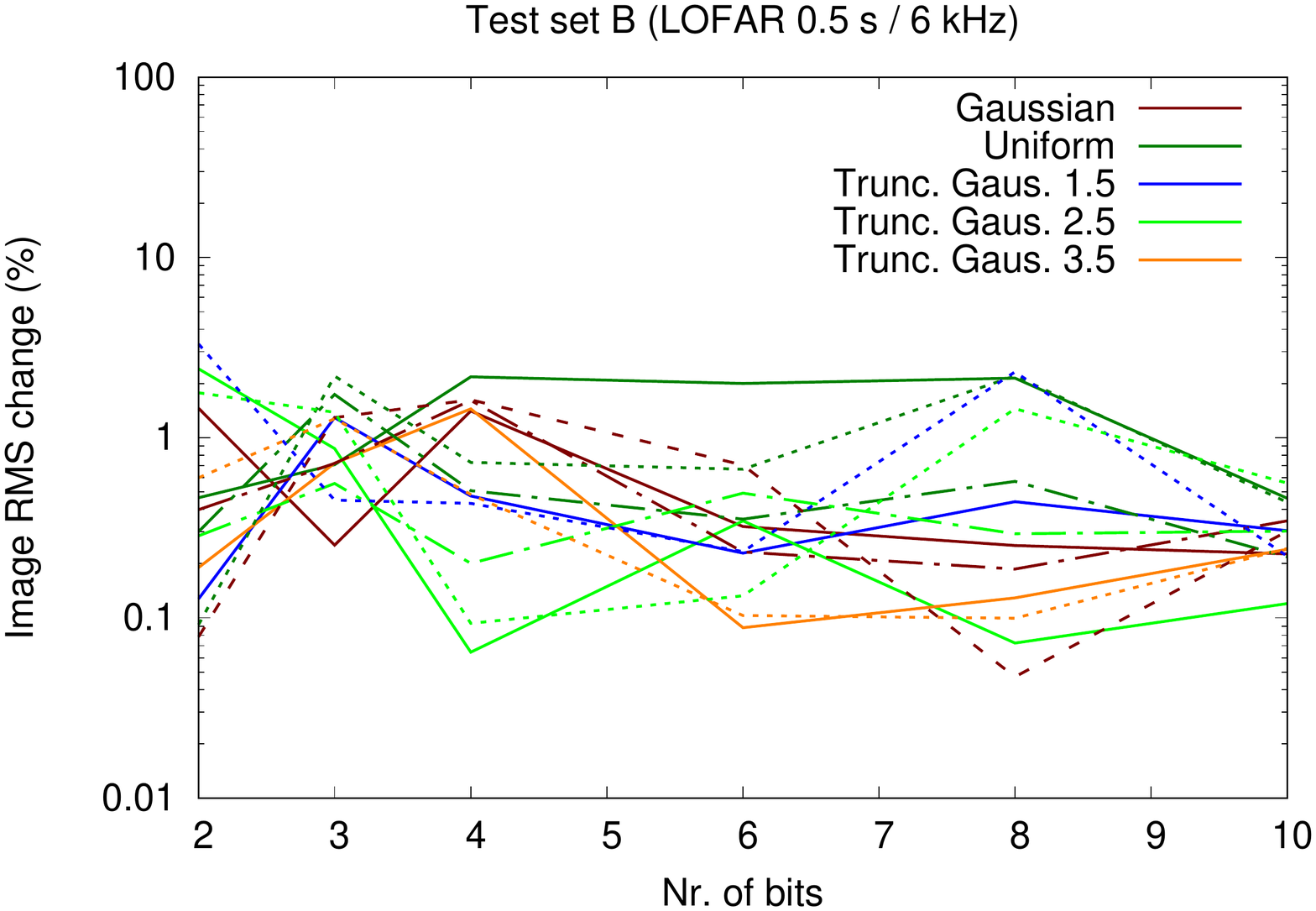}\\%
\hspace*{-6mm}\includegraphics[height=7.5cm]{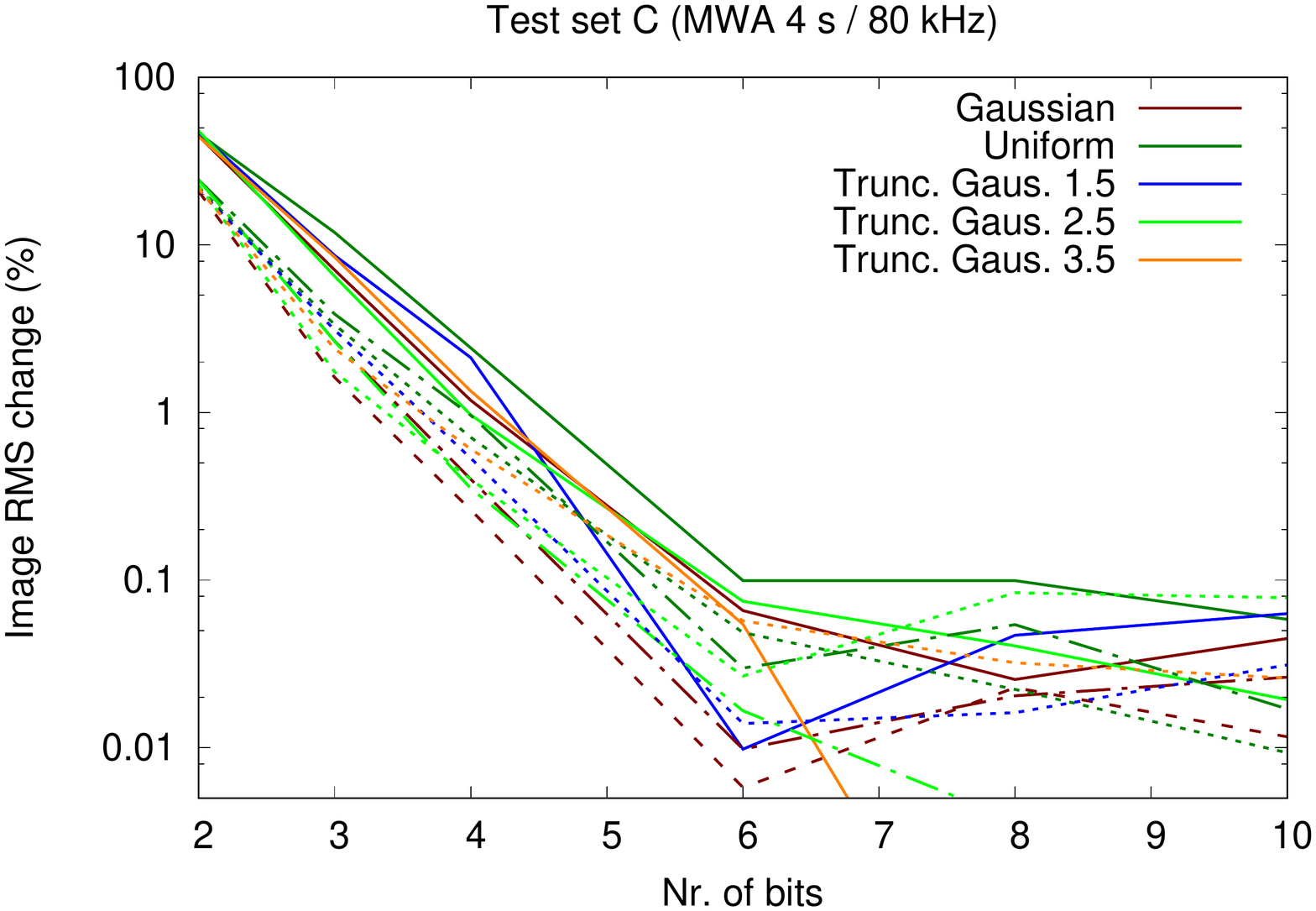}%
\hspace*{-16mm}\includegraphics[height=7.5cm]{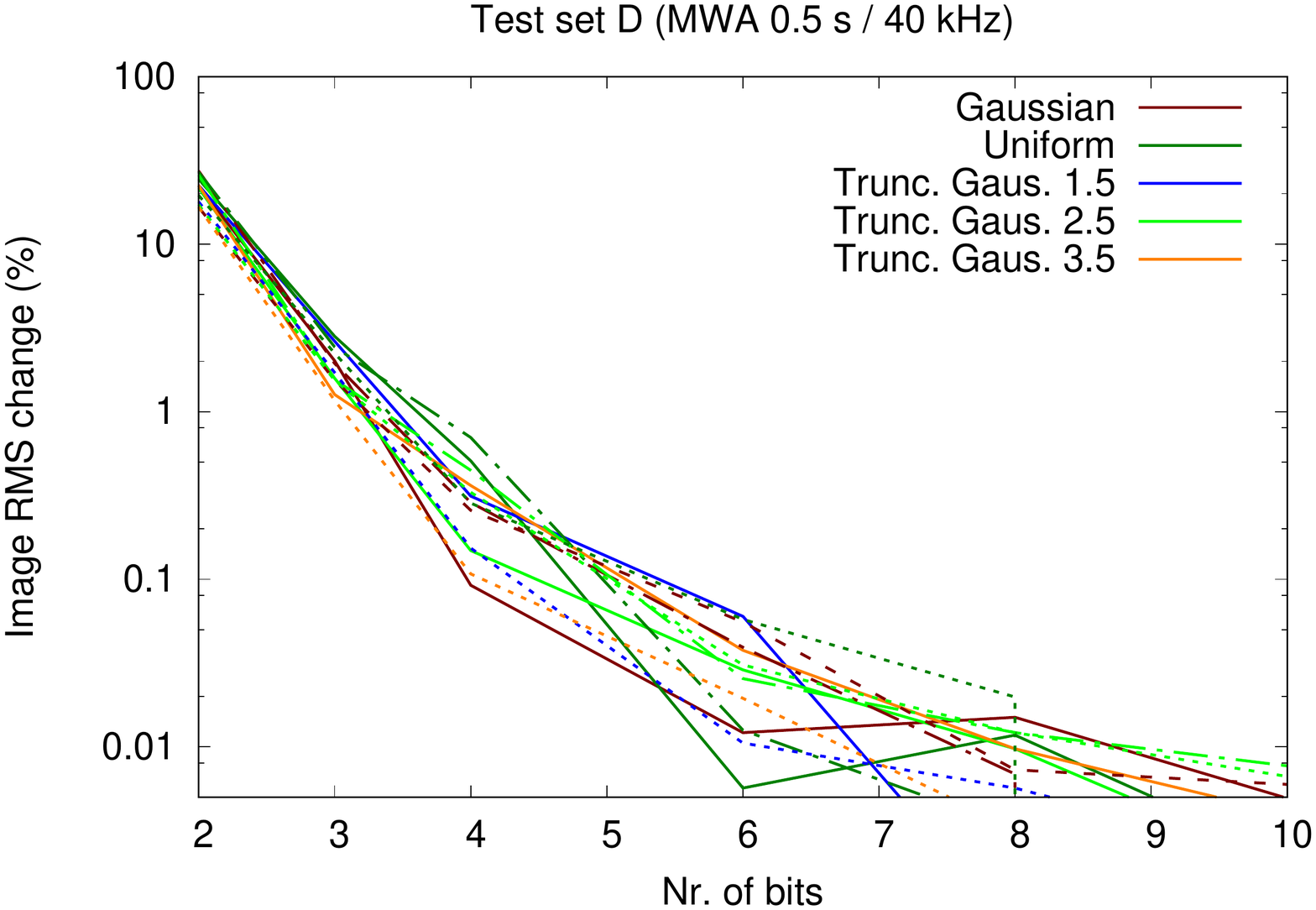}%
\caption{Change in imaging noise as an effect of compression, measured relatively to the uncompressed image noise in an empty region of the images. The measurements were performed for different quantization schemes and test sets. }
\label{fig:result-absnoise}%
\end{figure*}

Test set A is a LOFAR measurement set of the 3C~196 field, and is RFI flagged and compressed with one of the described methods. Subsequently, calibration and subtraction of 3C~196 is performed on compressed data. Calibration is performed using the Mitchcal tool introduced by \citet{offringa-2016}. The corrected visibilities are written without compression, so that the quantization noise is only added once. These uncompressed visibilities are then imaged. Because 3C~196 is a strong source at these frequencies ($\sim$100 Jy) while the imaging noise is only a few mJy, the set also tests compression of visibilities for imaging with high dynamic range.

Test set B is a LOFAR measurement set of the 3C~196 field with high time and frequency resolutions. It is RFI flagged before compression. After compression, it is averaged 8x in frequency and 16x in time, calibrated using the LOFAR calibration tool \textsc{ndppp} (Dijkema et al., in prep.) and imaged.

Test set C is an MWA measurement set targeting the Vela and Puppis A supernova remnants. It has been calibrated and averaged before compression. Calibration was performed by transferring solutions using CASA. The imaging is performed on the compressed visibilities. It has a relatively low time and frequency resolution, which is typical in MWA data processing to decrease the data volume. This set has a high SNR, in particular on the smaller baselines, because some of these see the Galactic plane.

Test set D is an MWA measurement set of calibrator Hydra A. Compression tests are performed on the original MWA time and frequency resolutions outputted by the MWA correlator. The compressed data is calibrated using Mitchcal, and the corrected visibilities are again stored with compression. This implies the compression noise is added twice.

All imaging is performed with \textsc{wsclean} \citep{offringa-wsclean-2014}, and all sets are imaged using uniform weighting. In uniform weighting, more weight is applied to the long baselines. Therefore, compression errors on the long baselines will dominate the compression errors in image space. This is a realistic use-case for data from these telescopes.

\subsection{Compression accuracy}

Fig.~\ref{fig:result-diffnoise} shows the differential imaging error (reference dirty image minus dirty image using compression) as a function of the number of bits used in compression, for different normalization and quantization techniques. Continuous lines are normalized using AF-normalization, dashed lines using RF-normalization and dot-dashed lines using row normalization. The colour indicates the quantization distribution. The horizontal grey line indicated the Stokes V noise, which is similar to the system noise. Hence, when the imaging error is well below this noise level, it will have an insignificant effect.

The Gaussian and 3.5$\sigma$ truncated Gaussian quantizations produce relatively larger errors compared to other quantizations if more than approximately 8 bits are used in compression. An intuitive explanation for this is that at higher bit levels, these distributions reserve more quantization values for outlying values. The visibilities are scaled such that the maximum visibility value matches the maximum quantization value. Because the number of visibilities per time block are limited, visibilities rarely have outlying values, and therefore do not match the quantization distribution.

With the exception of the Gaussian and 3.5$\sigma$-truncated Gaussian quantizations schemes, methods that perform relatively well with few bits also perform relatively well with more bits, which allows one to generalize the results of one bit-rate to other bit rates. Table~\ref{tbl:8bitresults} shows the 8-bit compression imaging errors for the three normalization methods and for the four sets. Additionally, it shows the average over the normalized sets for each unique combination of normalization and quantization method, as well as the average over the normalization methods and quantization methods. In order to make each test set equally important in the averaging, normalization is performed by dividing the results of a particular test set by the average result of that test set, and multiplying it by the average error over all test sets. Confidence intervals are given at 2$\sigma$ boundaries. Confidence intervals of average values are calculated by assigning each measurement a standard error equal to the standard deviation of the measurements that are averaged over, and these errors are propagated to the average value.

The results show that the combination of row-normalization with 1.5$\sigma$-truncated Gaussian quantization achieves the lowest error (153 $\pm$ 56 $\mu$Jy). The combination with second-lowest error is row-normalization with uniformly distributed quantization (177 $\pm$ 91$\mu$Jy). This is the ``AIPS compression'' technique, but with 8-bits instead of 16-bits quantization, and with dithering. As was shown in Fig.~\ref{fig:normalization-and-rfi-example}, per-row normalization is not robust and performs badly in the case of RFI or when there are other reasons for large differences between channels. The combination with lowest error excluding row-normalization, is AF-normalization with a 2.5$\sigma$-Gaussian quantization distribution (184 $\pm$ 31 $\mu$Jy).

When taking the average over the quantization methods, AF-normalization performs slightly but insignificantly better on average of the three normalization methods tested. When comparing the average error of the quantization methods by averaging over the normalization methods, the 1.5$\sigma$-truncated and 2.5$\sigma$-truncated Gaussian quantizations achieve the overall lowest errors. The Gaussian and 3.5$\sigma$-truncated Gaussian quantizations produce significantly larger errors than the other three methods.

In Fig.~\ref{fig:result-diffnoise}, for test sets A, B and C it is clear that the compression error decreases by a constant factor for each bit that is added. The fact that the decrease in error is larger at the very low bit-rates of 2--4 bits (which makes the error as a function of bit-rate slightly steeper in Fig.~\ref{fig:result-diffnoise}), is because one encoding quantity is reserved for the special value NaN. The error ratio of 8-bit compression over 16-bit compression is 258, 246 and 247 for set sets A, B and C respectively. Therefore, every added bit decreases the error by approximately a factor of 2. For test set D, the same 8-bit/16-bit ratio is only 5.8. This is likely because the compression error is not the dominating error in the imaging of this set. This set has more than 10 times more visibilities than the  other sets, and the large number of visibilities will increase the effect of quantization noise during the calibration and imaging due to the use of 32-bit floats in those operations. The small numerical changes caused by the compression will trigger a different instantiation of quantization noise during the imaging, and this dominates the error. Because the error is not decreased when going from 12 to 16 bits, the implication is that storing more than 12 bits is uneconomical for this test set, because the imaging error is already dominated by the numerical precision of the imaging.

So far, results were based on the error, i.e., the difference between the original dirty image and the dirty image after compression. To analyse the actual implications of compression after deconvolution, Fig.~\ref{fig:result-absnoise} shows the fractional noise changes calculated from cleaned Stokes-I images. Noise values are measured in a relatively empty square near the centre of the images. These plots mostly show the same trends as Fig.~\ref{fig:result-diffnoise}. For test set B, increasing the number of bits does not improve the sensitivity of the deconvolved image: even at the lowest bit-rate, the deconvolved image is dominated by system noise.

\begin{figure*}
\centering
\hspace*{-5mm}\includegraphics[height=6cm]{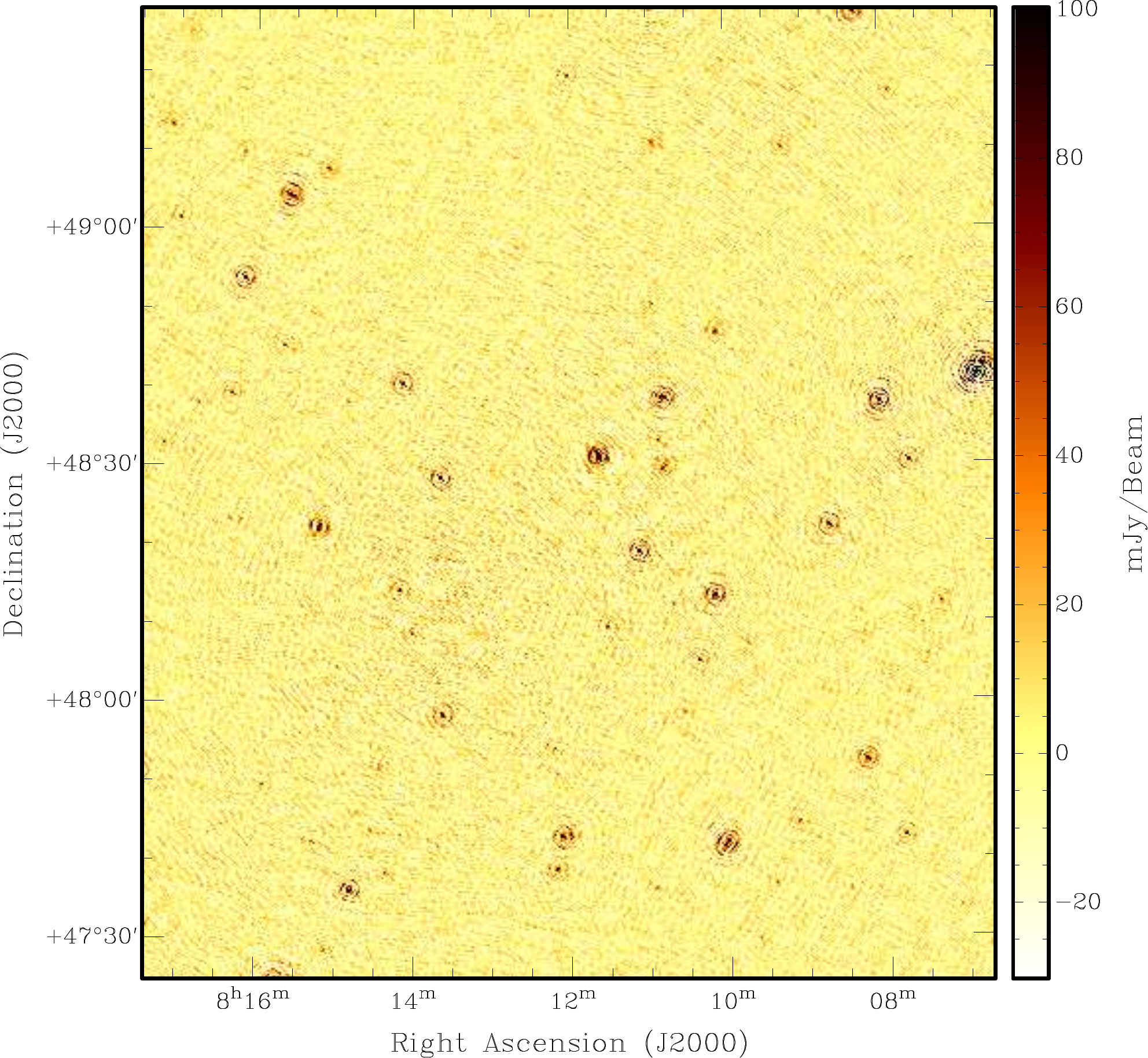}%
\hspace*{0mm}\includegraphics[height=6cm]{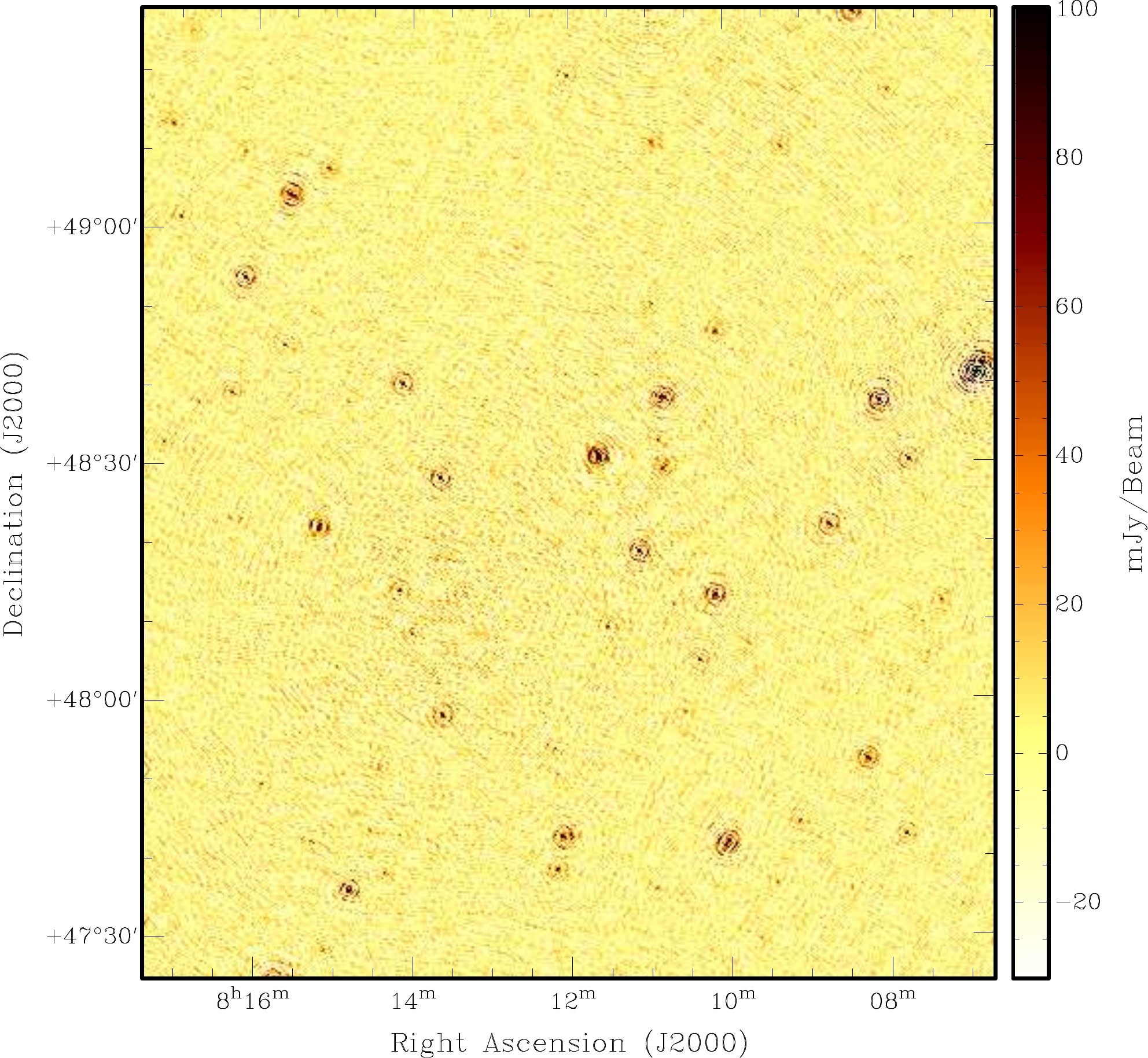}%
\hspace*{0mm}\includegraphics[height=6cm]{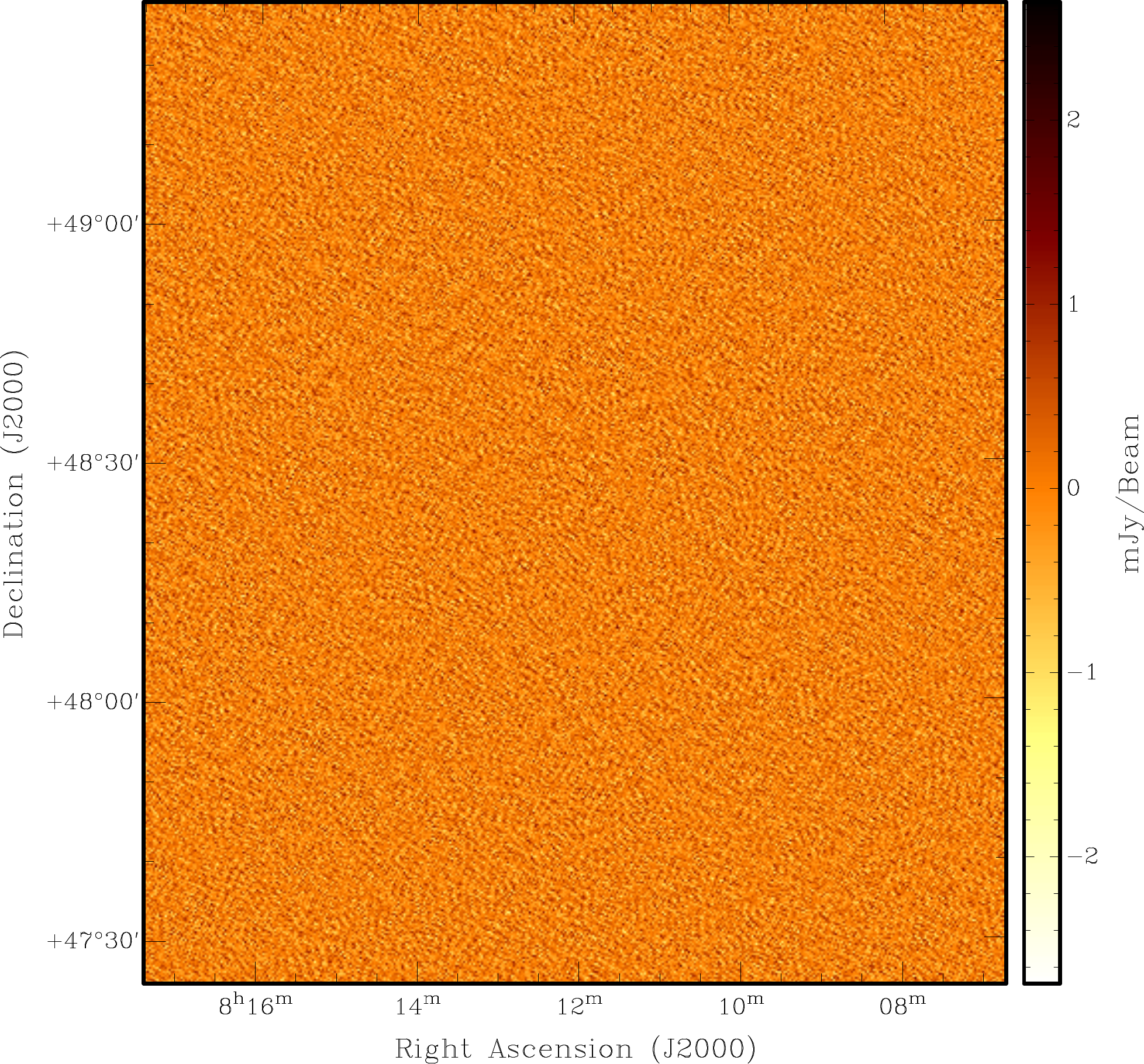}%
\caption{Noise added by 8-bit compression using LOFAR test set A. Left image: Result of calibration, 3C 196 subtraction and imaging without compression. Centre image: Same, but before processing the visibilities were compressed using the 8-bit quantization scheme (4$\times$ compression) with the 2.5$\sigma$-truncated Gaussian distribution. Right image: Difference between left and centre images -- note the different colour scheme. Its RMS is 430 $\mu$Jy. The added noise is unstructured and well below the system noise, and the left and centre images are visually indistinguishable.}
\label{fig:result-imaging-8bits}%
\end{figure*}

\begin{figure*}
\centering
\hspace*{-5mm}\includegraphics[height=6cm]{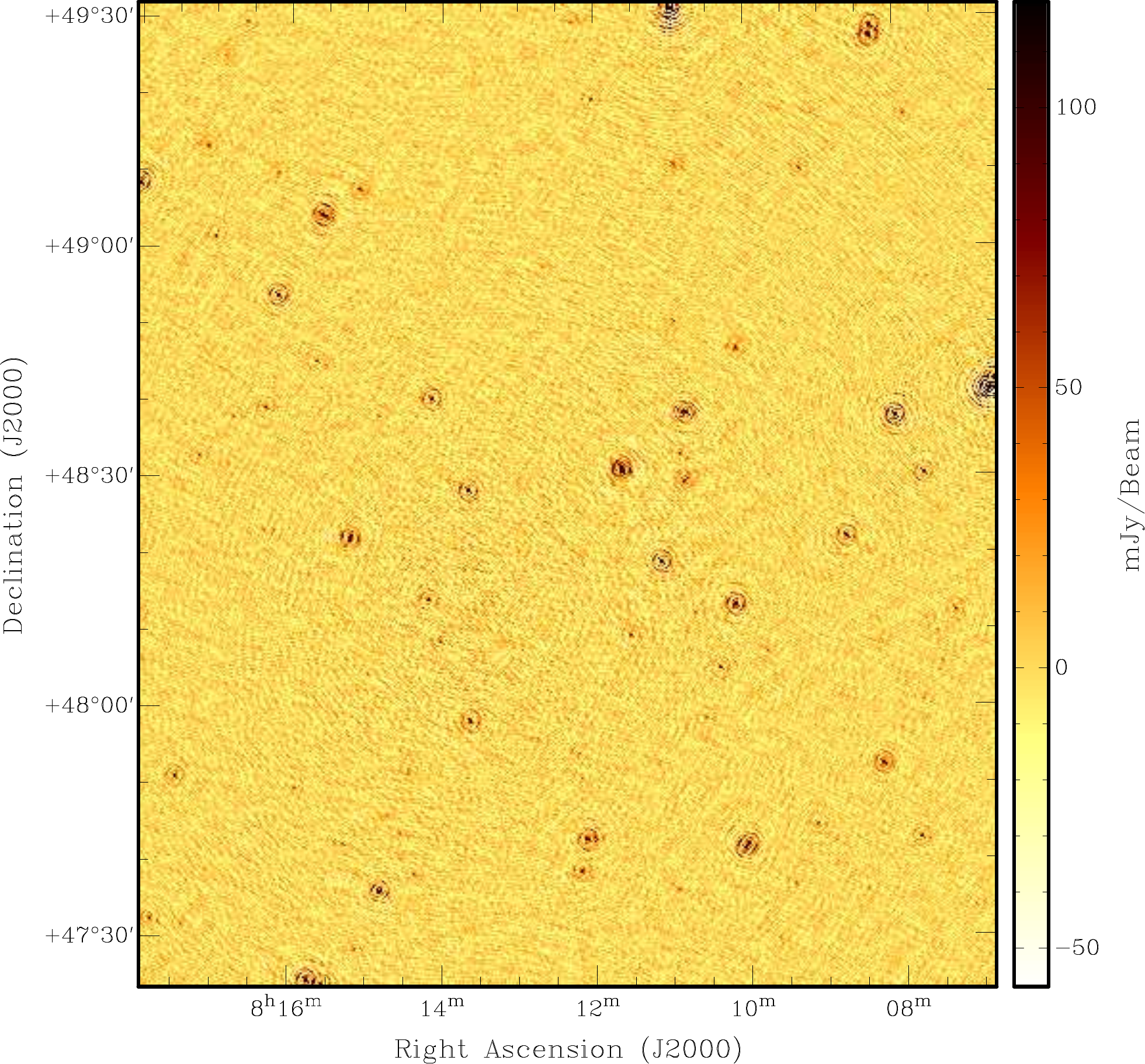}%
\hspace*{0mm}\includegraphics[height=6cm]{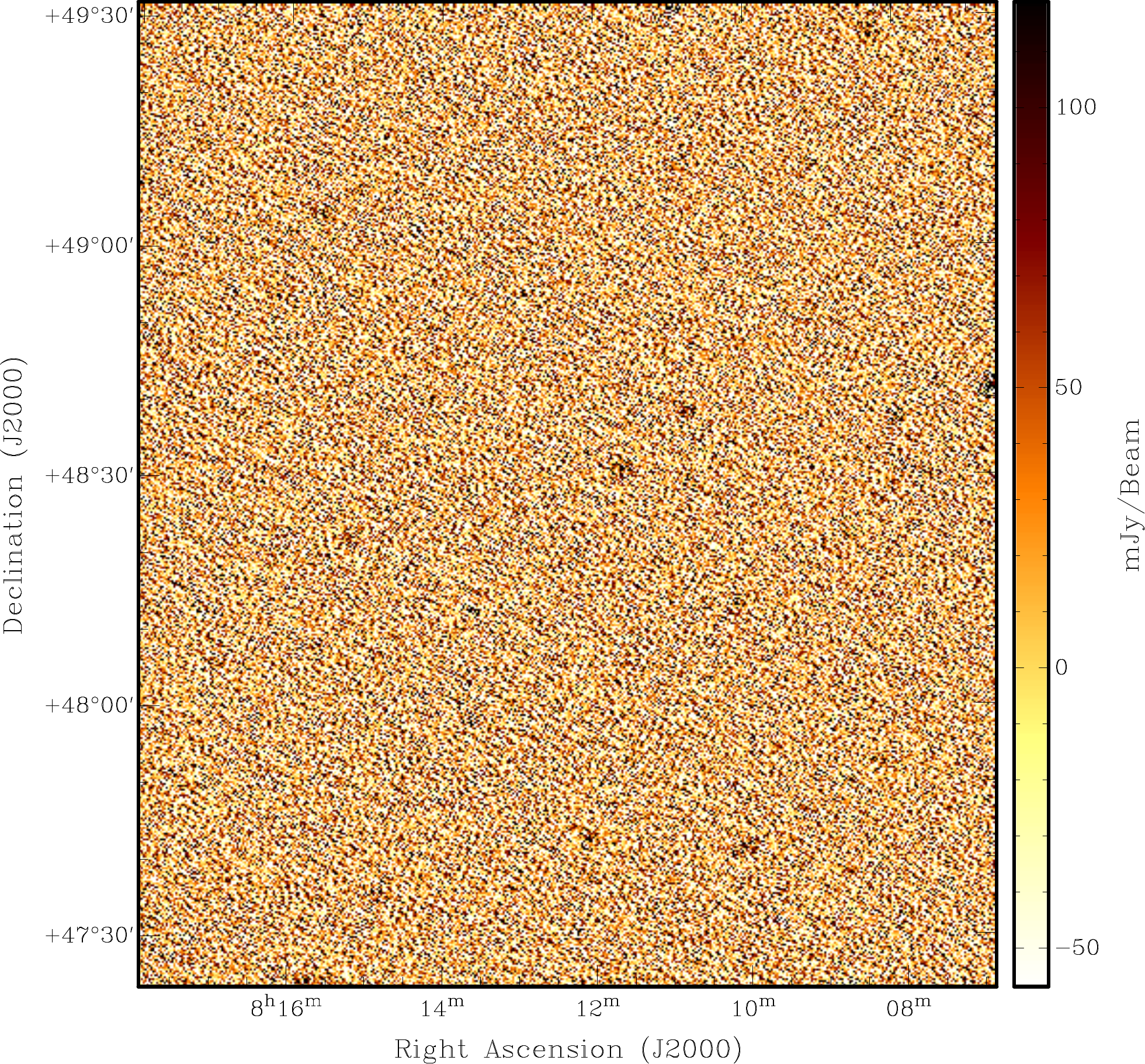}%
\hspace*{0mm}\includegraphics[height=6cm]{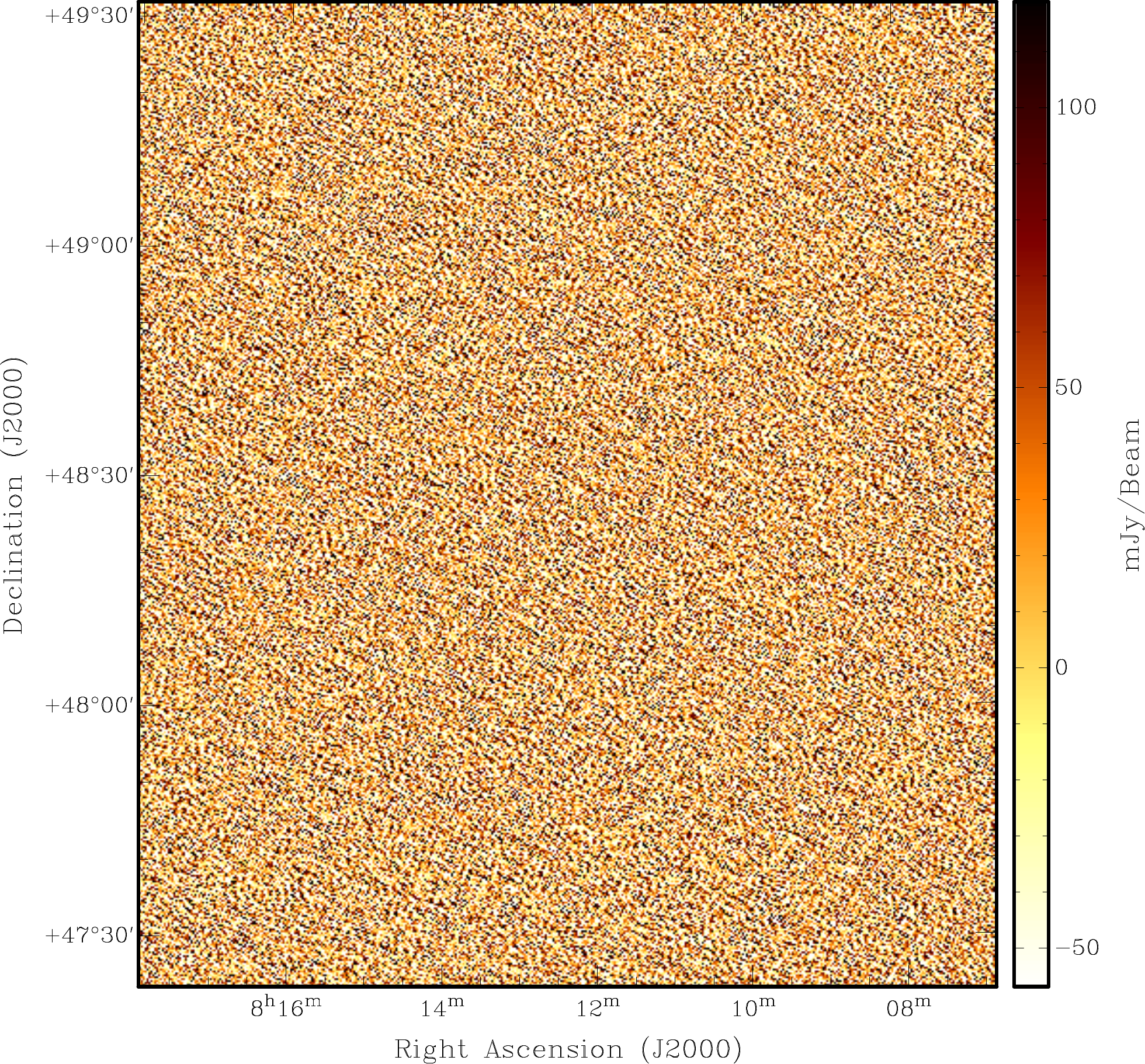}%
\caption{Noise added by 2-bit compression using LOFAR test set A. Left image: Results of calibration, 3C 196 subtraction and imaging without compression. Centre image: Same, but before processing the visibilities were compressed using the 2-bit quantization scheme (16$\times$ compression) with the 2.5$\sigma$-truncated Gaussian distribution. Right image: Difference between left and centre images. While the added compression noise dominates the noise in the image, the compression has not affected the sources systematically and the added noise is unstructured.}
\label{fig:result-imaging-2bits}%
\end{figure*}

\subsection{Properties of compression noise}
Besides producing low average errors, a visibility compression technique should also not add structural noise to an image, and should not change the flux of sources. To show that the compression technique indeed behaves properly, compression examples are shown in Figs.~\ref{fig:result-imaging-8bits} and \ref{fig:result-imaging-2bits}. Both figures display the original and compressed images and their difference, and are made from test set A.

In Fig.~\ref{fig:result-imaging-8bits}, an 8-bit quantization scheme is used with the 2.5$\sigma$-truncated Gaussian distribution and AF-normalization, such that the compression noise is lower than the image noise. The original and compressed images can not be distinguished by eye. The difference shows unstructured noise. The noise distribution of the image follows closely a Gaussian distribution.

In Fig.~\ref{fig:result-imaging-2bits}, a 2-bit quantization scheme is used with the 2.5$\sigma$-truncated Gaussian distribution and AF-normalization. As can be seen in the compressed image, the noise is significantly increased due to the low bit-rate and reasonably high SNR of the observation. The difference image again does not show any structure or bias at the position of strong sources. The noise distribution of the difference image follows again a Gaussian distribution.

From these results, it is clear that the quantization noise behaves like normal system noise in the image plane. While test sets B, C and D are short observations, test set A is a 6 hour observation. Because the noise in test set A does not show systematic behaviour, one can conclude that compression noise averages down like system noise, and remains uncorrelated in longer observations.

\subsection{Minimum bit-rates}
Using Fig.~\ref{fig:result-diffnoise} and \ref{fig:result-absnoise}, a minimum acceptable bit-rate can be derived by defining an acceptable imaging error. An acceptable imaging error might depend on the application: In a confusion-limited survey such as GLEAM \citep{wayth-2015-gleam,hurley-walker-gleam}, a compression noise addition at 10\% of the system noise is likely acceptable, since this will still be well below the imaging noise level. However, in system-noise-limited spectral-line work or in Epoch of Reionization experiments, a 10\% noise increase might not be acceptable. 

\begin{table} \caption{Minimum bit-rate per data set, given four different requirements. The second and third columns provide the minimum bit-rate for a differential error RMS < 10\% and 1\% of Stokes V RMS, respectively. Columns four and five provide the minimum bit-rate for an absolute noise change < 10\% and 1\%, respectively.}\label{tbl:minimum-bit-rates}\begin{center} \begin{tabular}{|l|c|c|c|c|} %
\hline
Set & $\Delta\sigma$<$0.1 \sigma_V$ & $\Delta\sigma$<$0.01 \sigma_V$ & $\sigma_c$<$1.1 \sigma_I $ & $\sigma_c$<$1.01 \sigma_I$ \\
\hline
A & 7 & 10 & 3 &  5 \\
B & 3 &  5 & 2 & ---\\
C & 5 &  8 & 3 &  4 \\
D & 3 &  6 & 2 &  3 \\
\hline
\end{tabular}\end{center}\end{table}

Minimum bit-rates are now calculated for four different requirements: for requirements i) and ii), the dirty image error RMS must be smaller than 10\% and 1\% of the system noise, respectively; for requirements iii) and iv), the deconvolved image noise level must change by less than 10\% and 1\%, respectively. Table~\ref{tbl:minimum-bit-rates} lists the results. With 10-bit compression, every test set can be imaged with less than 1\% error. Since a 1\% imaging error with a smaller imaging noise increase is probably acceptable in any project, it is likely that 10-bit compression is acceptable for all LOFAR observations, and 8-bit compression for all MWA observations.

Test sets B and D are at higher frequency and time resolution than A and C, and their SNR is thus lower. Therefore, they require fewer bits to achieve the same error. These sets can be compressed using 6-bit (LOFAR) and 5-bit (MWA) compression with less than 1\% error. Test sets B and D are compressed at the correlator output resolution, and this is in most cases the resolution at which the observations are archived. Hence, assuming the results can be generalized to other observations, LOFAR and MWA archival data can be compressed with an insignificant change in image quality using 5 and 6-bit compression, respectively.

\begin{figure}[htbp]
\centering
\hspace*{-11mm}\includegraphics[height=8cm]{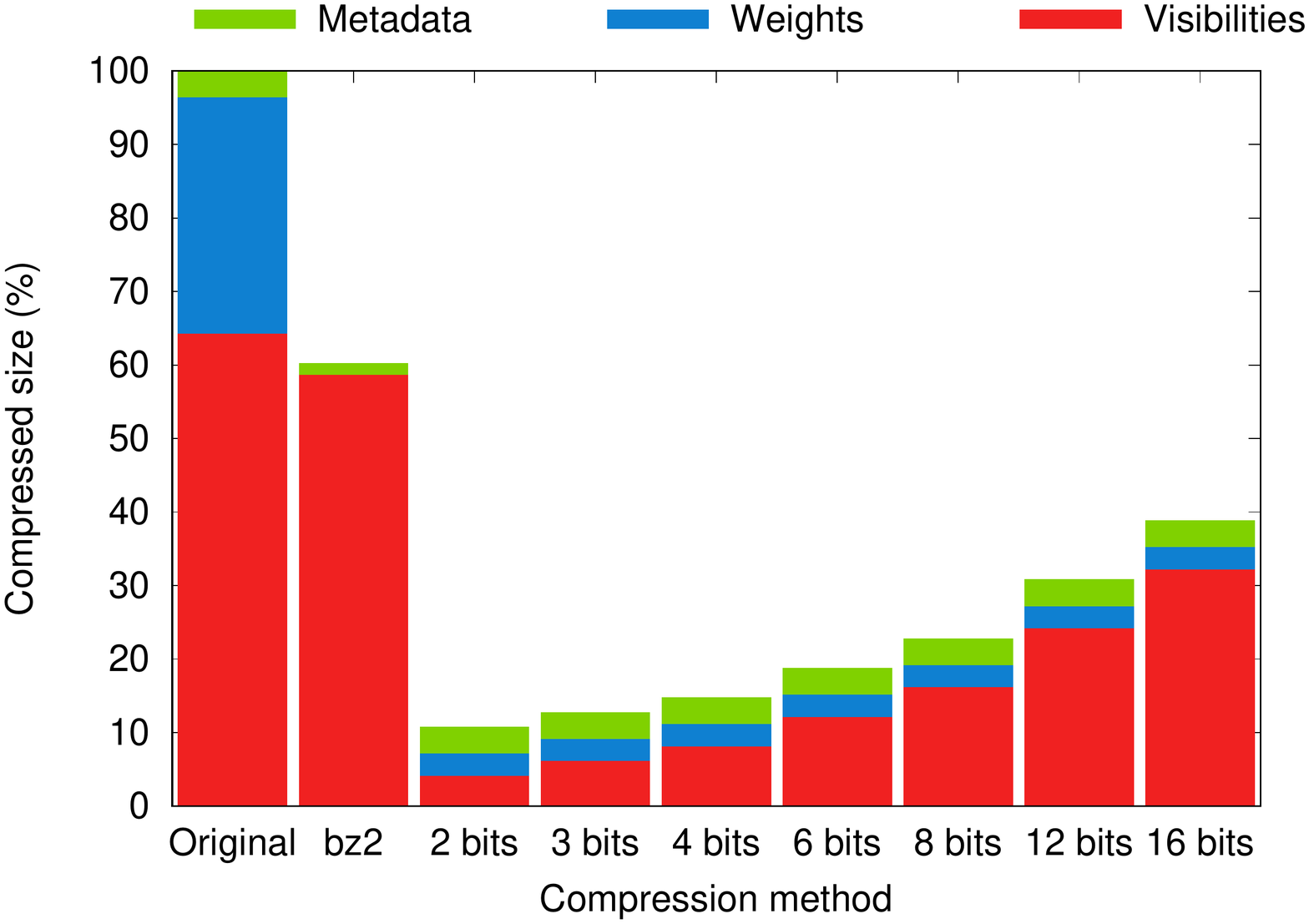}%
\caption{Comparison of data volume using test set B. The ``original'' case is not compressed, and ``bz2'' is compressed using \textsc{bzip2} compression. For the other cases, the weights are compressed to 12 bits and the visibilities are compressed to the value on the x-axis with one of the quantization methods described in this paper. }
\label{fig:volume-plot}%
\end{figure}
\subsection{Measurement set compression}
A Casacore measurement set can be decomposed into visibility data, weights and observational metadata\footnote{Here, observational metadata refers to other data inside the measurement set, such as antenna indices and timestamps. It is not to be confused with the normalization factors, which are here considered to be part of the visibility data volume.}. The measurement set definition allows two ways of storing weights: a single weight per table row that applies to all visibilities of that row; and a weight per visibility. In the latter case, the weights are stored in a column named ``WEIGHT\_SPECTRUM''. For each complex visibility consisting of two 32-bit floats, a single 32-bit float weight is stored. Therefore, the weights make up about one third of the volume of a data set. LOFAR and MWA observations require a weight per visibility, in particular because RFI can cause the weights of visibilities within a row to vary, and sensitivity will be lost if per-visibility weights are not stored and used in the processing. Another reason for per-visibility weights is to reduce the weight of the edges of a (sub)band with increased noise. Without per-visibility weights, these edges are not down-weighted and reduce the sensitivity.

When the DATA column is compressed by a factor of a few, it becomes relevant to compress the weights too. Fortunately, this is much less complicated. Normally, four linearly-polarized correlations are stored inside LOFAR measurement sets. The weights of visibilities for different polarizations are always the same. The measurement set format does not make use of this fact, and stores four equal weights in the case of LOFAR. By simply writing only one weight for the four polarized visibilities, the weight volume is reduced by a factor of 4. To increase the weight compression further, one can use the fact that quantizing the weights is much less problematic than quantizing visibilities: while changing the weights might reduce the sensitivity, even if the weights are always truncated, they will not cause a bias. Dithering is therefore not required. Weights are also not noisy, and can not have extreme outliers, making them suitable for linear quantization. I found that linearly quantizing the weights to 12 bits with row normalization does not influence the sensitivity noticeably ($\ll 1\%$) in the four test sets. The weights can probably be quantized with fewer bits, but extensive testing is limited in this work to compression of visibility data, which is the most pressing issue. Tests with different bit-rates for the weights were not performed. With 12-bit weights and one weight for every four differently-polarized visibilities, the weights are compressed by a factor of 10, and are no longer the dominating contribution to the size of compressed measurement sets.

Fig.~\ref{fig:volume-plot} shows the size of the data, weights and metadata of a measurement set that contains the data for test set B. The sizes before and after compression relative to the uncompressed size of the full measurement set are shown. The size of the measurement set after compression with the \textsc{bzip2} algorithm \citep{burrows-wheeler-1994} is also shown for comparison. With \textsc{bzip2}, the visibilities are hardly compressed. However, because weights do not contain noise and are repetitive, the weights can be compressed to less than 1\% of its original size. When 2 or 3 bit compression is used for the visibilities, the weights and metadata dominate the measurement set size.

\subsection{Computational performance} \label{sec:performance}

\begin{table} \caption{Wall-clock timings of several operations on test set D (65 GB) on a server with a 40-core Intel Xeon CPU E5-2660 @ 2.20GHz and 4 x 3TB RAID disks. Values are averages of 5 runs.}\label{tbl:performance} \begin{center} \begin{tabular}{lr} %
\hline
Reading the uncompressed set & 277 s \\
Copying the uncompressed set & 679 s \\
Compress visibilities with 8-bit / AF-normalization & 387 s \\
Compress visibilities with 2-bit / AF-normalization & 356 s \\
Compress visibilities with 8-bit / RF-normalization & 374 s \\
Compress visibilities with 2-bit / RF-normalization & 351 s \\
Compress visibilities with 8-bit / row normalization & 378 s \\
Compress visibilities with 2-bit / row normalization & 332 s \\
Compress weights to 12 bits & 285 s \\
Imaging of uncompressed set & 339 s \\ 
Imaging of 8-bit compressed set & 236 s \\
Imaging of 2-bit compressed set & 198 s\\
\hline
\end{tabular}
\end{center}
\end{table}

Table~\ref{tbl:performance} lists the wall-clock time of several operations. Each entry is the average of five runs. For the measurements, a high-end server with a 40-core Intel Xeon E5-2660 running at 2.20~GHz was used. For storage, a RAID disk with 4 x 3TB spinning disks was used. Since the ratio between compression and writing will vary strongly depending on the speed of the system, these results only provide an indicative impression of the speed of the compression.

Compressing the visibilities consists of reading the old data column and writing a new table column into the measurement set with the compressed visibilities. To replace an existing column with a compressed column, it is normally also required to replace flagged visibilities with NaN values. When performed as a separate step, this takes approximately as much time as compressing the visibilities. This is not included in the measurements. In a pipeline that writes out compressed visibilities, this step can be done on the fly. Once the compressed column has been written, the old uncompressed data column can be removed from the data set. However, the \textsc{casacore} storage managers do not release the space of a removed data column, and the measurement set will therefore not decrease in size. In order for the space to be released, the whole measurement set needs to be rewritten. Performing the full reordering is not included in these measurements. The reordering can be avoided by writing out compressed visibilities in a pipeline or by the correlator instead of replacing the data column of an existing measurement set.

The results show that compression speeds up processing of imaging. A considerable cost in compressing the visibilities is reading the uncompressed data, which can be avoided by integrating compression in a pipeline. Using a lower bit-rate decreases the wall-clock time of the operations, because it requires less reading or writing.

\section{Discussion} \label{sec:discussion}
It was shown that Dysco works best in cases with low SNR, where it can achieve compression factors of 5 to 6, or approximately twice as much if it is acceptable to increase the system temperature by 10\%. Since all four test sets are observations of bright targets, these test sets are approximately worst case situations. Most target fields will be more quiet, and will therefore compress with larger accuracy. While a few brighter targets exist (\object{Cygnus A} and \object{Cassiopeia A} in particular), these targets are often limited by dynamic range issues during deconvolution. It is therefore likely that an increase in system noise will not degrade the imaging results. In the exceptional situation of very bright targets and an extremely dynamic range, it is advisable to use no compression or compression with larger bit-rates.

Only LOFAR HBA and MWA sets were tested. Both tested telescopes operate at low frequency. It is likely that Dysco is effective for other telescopes, but further experiments are required to determine acceptable bit-rates. The LOFAR LBA measurements generally have lower SNR compared to LOFAR HBA, hence using the same compression settings for LBA as for HBA will be at least as accurate. In the current SKA design, the station size of SKA-low stations will be similar to the size of LOFAR stations and correlator resolutions will be at least as high as for LOFAR. It can therefore be assumed that the compression of SKA-low observations is equally affective as for LOFAR. Very long baseline interferometry (VLBI) observations are typically correlated at high time and frequency resolution, and produces noise-dominated data. Compression is therefore likely to work well on VLBI observations.

Averaging in time or frequency affects the field of view because of time and frequency smearing. It also removes information that might be useful for high-resolution spectral work or calibration. The compression methods introduced here do not remove the high-resolution information. Due to the dependence on the SNR, Dysco will in particular be useful for compressing raw correlation data. Because observations are often archived at high resolution, this can significantly lower the demands for archive space. During data processing, data from the archive is often averaged before calibration and imaging to decrease the computational requirements (e.g. \citealt{vanweeren-factor-2016, hurley-walker-gleam}). During this step, the averaged data can be written without compression. With the implementation resulting from this work, this can be completely transparent to the astronomer, because the compressed measurement set can be handled as a normal measurement set.

In some cases it might be helpful to compress averaged and/or calibrated data, for example to work with a limited amount of disk space or reduce IO overhead or network transfer. Because of the higher SNR, this requires more quantization bits, but compression factors of 3 and 4 can still be achieved for LOFAR and MWA, respectively.

The result that AF-normalization is on average more accurate than RF-normalization is somewhat counter-intuitive, since RF-normalization decomposes the data with more factors and has therefore more freedom to optimize the normalization. In fact, given the factors for AF-normalization, an RF-normalization can be found that performs equally well by setting each row factor to the multiplication of the two relevant antenna factors. Hence, RF-normalization never has to perform worse. The reason that this does occur is that, as discussed in \S\ref{sec:implementation}, the AF-normalization algorithm decomposes the data in a different order: first it normalizes the channel standard deviations, then it finds the antenna factors, and finally it iteratively maximizes the channel factor or antenna factor that increases the sum of absolute value the most, until convergence. In the RF-normalization algorithm, the standard deviations of the channels are also first normalized, but then each row is maximized. Afterwards, each channel is maximized. The latter algorithm is easier and computationally faster than the AF-normalization algorithm, but results evidently in less optimal factors. An algorithm that produces RF-normalization factors that is at least as accurate as AF-normalization, is to calculate the factors by starting from the AF-normalization factors, and then maximize each row. The computational cost of this is relatively low compared to astronomical operations such as calibration or imaging, so the extra computational cost is likely not an issue. Given that the noise variance of observations can physically be described by the AF-normalization factors, on noise-dominated sets it is likely that the extra normalization freedom of RF-normalization does not contribute much.

This work uses for the first time a non-linear quantization scheme on radio astronomical data, in which the quantization is matched to the expected distribution of the input data. In AF-normalization and RF-normalization it reduces the quantization error compared to linear quantization by 1.60 and 1.25 on average, respectively. Non-linear quantization is therefore indeed a small improvement over linear quantization for the encoding of radio data.

Our current implementation consists of a storage manager used by \textsc{casacore} and a stand-alone tool that can replace the column of an existing measurement set with a compressed column. As shown, compression works best on RFI flagged data, which is often performed by a preprocessing pipeline. For example, LOFAR uses \textsc{ndppp} and MWA uses \textsc{cotter}, two preprocessing pipelines that flag the data using the \textsc{aoflagger}, convert the format of the data to a standard measurement set, and can do several other tasks at the same time. Because raw data from these telescopes have large volumes, every extra read or write of the raw data is costly. Hence, the ideal place for compression is inside these pipelines, just before writing to disk. Pipelines can do so by requesting \textsc{casacore} to use the compressing storage manager for the relevant columns.

As was shown in \S\ref{sec:performance}, writing data with compression is faster than writing without compression. In certain processing steps, this might be an important consideration. In the Factor pipeline \citep{vanweeren-factor-2016}, disk input-output is a considerable cost, because the original data needs to be read, phase changed, averaged and rewritten to a measurement set for each facet direction. One possible way to improve this, is to compress the high-resolution data at the original phase direction. This will have an insignificant loss in accuracy due to the low SNR, yet decrease the IO-overhead considerably.

\subsection{Comparison with other implementations}
In this section, I will briefly compare the features of Dysco to the \textsc{aips} 16-bit compression technique, the FITS file compression technique and the \textsc{bitshuffle} compression technique.

\textsc{aips} uses row-normalization with uniform quantization, which I showed to have a high accuracy on average. As was shown, it performs not well in the case of RFI or high dynamic range. The bit-rate of 16 bits per input value allows about 4 orders of magnitude of dynamic range before catastrophic quantization occurs. Flagging the data before compression will mostly mitigate this problem. The RF and AF-normalization schemes introduced in this work also solve this issue. The lack of dithering in \textsc{aips} compression might lead to a systematic bias \citep{pence-2010-fits-dithering}. This is only an issue in the case of high SNR. In the case of low SNR, the intrinsic noise in the visibilities will mimic dithering. Of course, with \textsc{aips} the compression rate is limited to a factor of 2.

The \textsc{bitshuffle} compression technique also uses uniform quantization. Before quantization, values are rounded relative to the expected noise in a sample. The effect of this is somewhat comparable to the combination of normalization and quantization used in this work, which also throws away significance that is smaller than the noise.

The \textsc{bitshuffle} and FITS compression technique use more bits when the rounded values have higher entropy. Therefore, it compresses effectively with a variable bit-rate. This is effective for signal-dominated data with repeating values, but has no benefit for the compression of noise-dominated data with constant entropy.

The FITS compression technique uses exponential quantization. Thereby, similar to Gaussian quantization, the quantization error is smaller for values near zero. In exponential quantization, small values will be preserved with exponentially more precision than large values. This is useful for image data, where the relative error on each pixel is relevant, but is not optimal for visibilities. For those, the average error is a more relevant measure of the effects on the final science products, and since visibilities follow a Gaussian distribution -- or even a more uniform distribution after normalization -- quantizing with an exponential distribution is not optimal.

\subsection{Guideline for method selection}
While it is possible to measure the compression loss of observations and determine the acceptable compression method, this is not practical for regular observing of different targets with different correlation settings. Instead, in this section I will weigh the characteristics of the various compression configurations and suggest a rule for selecting the compression method that will not affect data.

Based on the accuracy, several combinations of quantization and normalization have been shown to perform well on all test sets. Per-row normalization with uniform quantization, similar to the method used by \textsc{aips} but including dithering and different bit-rates, is on average one of the most accurately performing methods. Moreover, it is the simplest method to implement. However, as was shown, the accuracy of compression with row normalization is sensitive to large dynamic ranges in the visibility, such as in the case of RFI (Fig.~\ref{fig:normalization-and-rfi-example}), and makes the compression of the different polarizations dependent on each other (Fig.~\ref{fig:normalization-examples-scatter}). The RF and AF normalizations are robust to these issues. On average, the AF normalization performs slightly better than the RF-normalization method, and the most accurate combination with AF-normalization is more accurate than the most accurate combination with RF-normalization. The difference between accuracy of the normalization methods with best quantization are however not significant.

AF-normalization is theoretically slightly more computational intensive during compression compared to RF-normalization and row-normalization, but this difference is small in practice. The methods require the same time to decompress. The quantization method does not change the computational performance of the methods. AF-normalization requires special treatment of auto-correlations, but requires less metadata to be stored ($N_\textrm{pol} (N_\textrm{ch} + N_\textrm{ant})$ vs. $N_\textrm{pol} (N_\textrm{ch} + N_\textrm{ant}^2)$ per timeblock). In large measurement sets with many channels, the metadata is insignificant. All in all, there is little reason to select one method over another.

Noise normalization performed by AF normalization will equalize the statistics of all visibilities. Therefore, the extra factors that RF normalization adds do not match the physical process by which the noise is formed. When the signal adds a dominant contribution to the visibilities, the RF-normalization can have a benefit, because with RF-normalization the visibilities of different baselines can be scaled independently. The results show that AF normalization combines better with centrally-dense quantization distributions, while RF normalization favours more uniform distributions. For sets with larger number of channels or antennas, the statistics of sets after normalization will be more accurately quantized by centrally-dense distributions.

For these reasons, I suggest to use the following guideline for compression of MWA and LOFAR observations, as well as future high-resolution SKA observations:
\begin{itemize}
\item[-] For the compression of high-resolution observations ($\Delta t \times \Delta \nu \le 0.5$s$\times 6$kHz for LOFAR and SKA, and $\Delta t \times \Delta \nu \le 0.5$s$\times 40$kHz for the MWA), use AF normalization with 2.5$\sigma$-truncated Gaussian quantization. As described, on average it is the most accurate method with the exception of the less stable row-normalization, it is robust and it matches the underlying physics. High-resolution LOFAR sets can be compressed with 5-bit quantization and MWA sets can be compressed with 6-bit quantization, respectively.
\item[-] Observations with a lower resolution -- and therefore higher SNR -- should be compressed using higher bit-rates. Up to a resolution of $\Delta t \times \Delta \nu \le 4$s$\times 36$kHz for LOFAR and SKA-low, and $\Delta t \times \Delta \nu \le 4$s$\times 80$kHz for the MWA, RF-normalization with 1.5$\sigma$-truncated Gaussian quantization should be used, with 10-bit compression for LOFAR and 8-bit compression for the MWA. On the lower-resolution sets A and C, this is on average the most accurate compression method. As was shown, it is also robust.
\end{itemize}
Compression with these parameters increases the image noise with less than 1\%.

\section{Possible improvements \& future work} \label{sec:futurework}
After the quantization of randomly distributed Gaussian noise with a matched quantizer, the compressed data will be independently-distributed noise with a uniform distribution that can not be compressed any further. Hence, in noise-dominated situations, using the distribution of the data and normalizing the data with AF-normalization, will lead in theory to the best compression rate. In the case of signal-dominated data, compressed data will be correlated and can be compressed further. To make use of such correlations, an encoding scheme such as Rice coding \citep{1971-rice-coding} or LZ77 compression \citep{ziv-1977-lz77} can be used to increase the compression factor. This makes the input-output of data somewhat more complex, because different timeblocks can have different sizes after Rice coding, and this requires extra administration to keep track of the place and size of each timeblock. The algorithm will also have to look for correlations in the time direction, while time is generally the slowest changing dimension. It has been shown to provide good results thought \citep{masui-2015-compression}, and a combination of these methods might provide a generic algorithm that performs near-optimal in noise-dominated as well as signal-dominated situations.

As was shown, metadata and weights are easily compressible. At low visibility bit-rates, these can have a significant size compared to the data. While it was shown that the weights can be compressed by a factor of 10, compression with \textsc{bzip2} can compress the weights by more than a factor of 100. Hence, a similar algorithm but one that can randomly and transparently access the weights inside a measurement set will increase the compression of the weights, compared to the method used in this work.

Baseline-dependent averaging can also reduce the size of observations. This can in theory be performed together with the compression method described in this paper. In sets that are not baseline-dependently averaged, the visibilities of long baselines receive more weight and will therefore dominate the compression error in image space. Smaller baselines can be stored with fewer bits before their inaccuracies becomes noticeable in image plane. With baseline-dependent averaging, the weight of each visibility will be more constant when using the uniform weighting scheme, and every visibility will contribute similarly to the compression error (in image plane). However, if averaging of the smaller baselines increases the SNR in those baselines too much, compression will be less accurate. In that case, the smaller baselines might dominate the compression error. The application of baseline-dependent averaging is limited for compressing archival data, because accurate RFI flagging require high time and frequency resolutions \citep{lofar-radio-environment} and baselines can only be averaged up to the calibration solution interval. For LOFAR, this interval can be as short as a few seconds to deal with the ionosphere in high-resolution maps \citep{vanweeren-factor-2016}.

Currently, the distribution of a particular visibility is parametrized by a single scaling factor. It is evident from the test results that this single parametrization is not always optimal; signal-dominated data sets are more accurately compressed by using a uniform distribution in the quantizer, while noise-dominated data sets are more accurately compressed by a truncated Gaussian distribution. Therefore, a possible increase in accuracy could be achieved by increasing the number of parameters of the distribution within each timeblock, for example by searching the optimal truncation value for each timeblock and storing these along with the scaling factors.

\section{Conclusions} \label{sec:conclusions}
The Dysco technique for compressing visibilities is suitable for radio observations. The noise added by this compression technique acts like normal system noise. The accuracy of the compression is depending on the signal-to-noise ratio of the data: noisy data can be compressed with a smaller loss of image quality. Data with typical correlator time and frequency resolutions can be compressed by a factor of 6.4 for LOFAR and 5.3 for MWA observations with less than 1\% added system noise in image plane. After averaging observations in time and frequency to the typical resolutions used in processing, a compression factor of 3.2 to 4 can be reached with less than 1\% added system noise in image plane. The technique is in particular well suited to reduce the archival space requirements.

So far, testing was performed only on low-frequency data from the MWA and LOFAR telescopes. The implementation is generic and can be applied to other telescopes. However, further experiments are required to determine acceptable bit-rates.

\begin{acknowledgements}
I acknowledge financial support from the European Research Council under ERC Advanced Grant LOFARCORE 339743.
\end{acknowledgements}

\label{lastpage}

\DeclareRobustCommand{\TUSSEN}[3]{#3}

\bibliographystyle{aa} 
\bibliography{references}

\begin{thebibliography}{28}
\expandafter\ifx\csname natexlab\endcsname\relax\def\natexlab#1{#1}\fi

\bibitem[{Burrows \& Wheeler(1994)}]{burrows-wheeler-1994}
Burrows, M. \& Wheeler, D.~J. 1994, Technical report 124, Digital Equipment
  Corporation, http://www.hpl.hp.com/techreports/Compaq-DEC/SRC-RR-124.html

\bibitem[{Favre {et~al.}(2011)Favre, Wootten, Remijan, Brouillet, Wilson,
  Despois, \& Baudry}]{evla-high-resolution-lines}
Favre, C., Wootten, H.~A., Remijan, A.~J., {et~al.} 2011, The Astrophysical
  Journal Letters, 739, L12

\bibitem[{Greisen(2016)}]{aips-fits-format-2016}
Greisen, E.~W. 2016, Tech. Rep., AIPS Memo 117,
  http://www.aips.nrao.edu/aipsmemo.html

\bibitem[{{\TUSSEN{Haarlem}{Van}{van}}~Haarlem
  {et~al.}(2013){\TUSSEN{Haarlem}{Van}{van}}~Haarlem, Wise, Gunst,
  {et~al.}}]{lofar-2013}
{\TUSSEN{Haarlem}{Van}{van}}~Haarlem, M.~P., Wise, M.~W., Gunst, A.~W.,
  {et~al.} 2013, A\&A, 556, A2

\bibitem[{Heywood {et~al.}(2016)Heywood, Bannister, Marvil,
  {et~al.}}]{heywood-2016-askap}
Heywood, I., Bannister, K.~W., Marvil, J., {et~al.} 2016, MNRAS, 457, 4160

\bibitem[{Hurley-Walker {et~al.}(submitted)}]{hurley-walker-gleam}
Hurley-Walker, N. {et~al.} submitted

\bibitem[{Jiwani {et~al.}(2013)Jiwani, Colegate, Razavi-Ghods, Hall, Padhi, \&
  bij~de Vaate}]{ska-station-config-2013}
Jiwani, A., Colegate, T., Razavi-Ghods, N., {et~al.} 2013, in press

\bibitem[{Kazemi {et~al.}(2013)Kazemi, Yatawatta, \&
  Zaroubi}]{kazemi-clustered-cal-2013}
Kazemi, S., Yatawatta, S., \& Zaroubi, S. 2013, Monthly Notices of the Royal
  Astronomical Society, 430, 1457

\bibitem[{Masui {et~al.}(2015)Masui, Amiri, Connor,
  {et~al.}}]{masui-2015-compression}
Masui, K., Amiri, M., Connor, L., {et~al.} 2015, Astronomy and Computing, 12,
  181

\bibitem[{Morabito {et~al.}(2014)Morabito, Oonk, Salgado,
  {et~al.}}]{morabito-2014-carbon-rrls}
Morabito, L.~K., Oonk, J. B.~R., Salgado, F., {et~al.} 2014, ApJL, 795, L33

\bibitem[{Offringa {et~al.}(2013)Offringa, de~Bruyn, Zaroubi,
  {et~al.}}]{lofar-radio-environment}
Offringa, A.~R., de~Bruyn, A.~G., Zaroubi, S., {et~al.} 2013, A\&A, 549

\bibitem[{Offringa {et~al.}(2014)Offringa, McKinley, Hurley-Walker,
  {et~al.}}]{offringa-wsclean-2014}
Offringa, A.~R., McKinley, B., Hurley-Walker, N., {et~al.} 2014, MNRAS, 444,
  606

\bibitem[{Offringa {et~al.}(2016)Offringa, Trott, Hurley-Walker,
  {et~al.}}]{offringa-2016}
Offringa, A.~R., Trott, C.~M., Hurley-Walker, N., {et~al.} 2016, MNRAS, 458,
  1057

\bibitem[{Pence {et~al.}(2010)Pence, White, \&
  Seaman}]{pence-2010-fits-dithering}
Pence, W.~D., White, R.~L., \& Seaman, R. 2010, Publications of the
  Astronomical Society of the Pacific, 122, 1065

\bibitem[{Perley {et~al.}(2011)Perley, Chandler, Butler, \&
  Wrobel}]{evla-perley-2011}
Perley, R.~A., Chandler, C.~J., Butler, B.~J., \& Wrobel, J.~M. 2011, ApJ
  Letters, 739, L1

\bibitem[{Rice \& Plaunt(1971)}]{1971-rice-coding}
Rice, R. \& Plaunt, J. 1971, IEEE Transactions on Communication Technology, 19,
  889

\bibitem[{Shannon \& Weaver(1949)}]{shannon-entropy-definition-1949}
Shannon, C.~E. \& Weaver, W. 1949, The Mathematical Theory of Communication
  ({Univ of Illinois Press})

\bibitem[{{Smirnov, O.~M.}(2011)}]{revisiting-me-ii}
{Smirnov, O.~M.} 2011, A\&A, 527, A107

\bibitem[{{Stappers, B. W.} {et~al.}(2011){Stappers, B. W.}, {Hessels, J. W.
  T.}, {Alexov, A.}, {Anderson, K.}, {Coenen, T.}, {Hassall, T.},
  {Karastergiou, A.}, {Kondratiev, V. I.}, {Kramer, M.}, {van Leeuwen, J.},
  {Mol, J. D.}, {Noutsos, A.}, {Romein, J. W.}, {Weltevrede, P.}, {Fender, R.},
  {Wijers, R. A. M. J.}, {B\"ahren, L.}, {Bell, M. E.}, {Broderick, J.}, {Daw,
  E. J.}, {Dhillon, V. S.}, {Eisl\"offel, J.}, {Falcke, H.}, {Griessmeier, J.},
  {Law, C.}, {Markoff, S.}, {Miller-Jones, J. C. A.}, {Scheers, B.}, {Spreeuw,
  H.}, {Swinbank, J.}, {ter Veen, S.}, {Wise, M. W.}, {Wucknitz, O.}, {Zarka,
  P.}, {Anderson, J.}, {Asgekar, A.}, {Avruch, I. M.}, {Beck, R.}, {Bennema,
  P.}, {Bentum, M. J.}, {Best, P.}, {Bregman, J.}, {Brentjens, M.}, {van de
  Brink, R. H.}, {Broekema, P. C.}, {Brouw, W. N.}, {Br\"uggen, M.}, {de Bruyn,
  A. G.}, {Butcher, H. R.}, {Ciardi, B.}, {Conway, J.}, {Dettmar, R.-J.}, {van
  Duin, A.}, {van Enst, J.}, {Garrett, M.}, {Gerbers, M.}, {Grit, T.}, {Gunst,
  A.}, {van Haarlem, M. P.}, {Hamaker, J. P.}, {Heald, G.}, {Hoeft, M.},
  {Holties, H.}, {Horneffer, A.}, {Koopmans, L. V. E.}, {Kuper, G.}, {Loose,
  M.}, {Maat, P.}, {McKay-Bukowski, D.}, {McKean, J. P.}, {Miley, G.},
  {Morganti, R.}, {Nijboer, R.}, {Noordam, J. E.}, {Norden, M.}, {Olofsson,
  H.}, {Pandey-Pommier, M.}, {Polatidis, A.}, {Reich, W.}, {R\"ottgering, H.},
  {Schoenmakers, A.}, {Sluman, J.}, {Smirnov, O.}, {Steinmetz, M.}, {Sterks, C.
  G. M.}, {Tagger, M.}, {Tang, Y.}, {Vermeulen, R.}, {Vermaas, N.}, {Vogt, C.},
  {de Vos, M.}, {Wijnholds, S. J.}, {Yatawatta, S.}, \& {Zensus,
  A.}}]{lofar-pulsars-and-transients-2011}
{Stappers, B. W.}, {Hessels, J. W. T.}, {Alexov, A.}, {et~al.} 2011, A\&A, 530,
  A80

\bibitem[{Thompson {et~al.}(2007)Thompson, Emerson, \&
  Schwab}]{quantization-efficiency-2007}
Thompson, A.~R., Emerson, D.~T., \& Schwab, F.~R. 2007, Radio Science, 42

\bibitem[{Thompson {et~al.}(2001)Thompson, Moran, \&
  Swenson}]{thompson-radio-interferometry}
Thompson, A.~R., Moran, J.~M., \& Swenson, G.~W. 2001, {Interferometry and
  Synthesis in Radio Astronomy, 2nd edition} (Wiley-Interscience)

\bibitem[{Tingay {et~al.}(2013)Tingay, Goeke, Bowman, Emrich, Ord, Mitchell,
  Morales, Booler, Crosse, Wayth, Lonsdale, Tremblay, Pallot, Colegate,
  Wicenec, Kudryavtseva, Arcus, Barnes, Bernardi, Briggs, Burns, Bunton,
  Cappallo, Corey, Deshpande, Desouza, Gaensler, Greenhill, Hall, Hazelton,
  Herne, Hewitt, Johnston-Hollitt, Kaplan, Kasper, Kincaid, Koenig,
  Kratzenberg, Lynch, Mckinley, Mcwhirter, Morgan, Oberoi, Pathikulangara,
  Prabu, Remillard, Rogers, Roshi, Salah, Sault, Udaya-Shankar, Schlagenhaufer,
  Srivani, Stevens, Subrahmanyan, Waterson, Webster, Whitney, Williams,
  Williams, \& Wyithe}]{mwa-2013-tingay}
Tingay, S.~J., Goeke, R., Bowman, J.~D., {et~al.} 2013, PASA, 30

\bibitem[{Van~Cappellen \& Bakker(2010)}]{apertif-2010}
Van~Cappellen, W. \& Bakker, L. 2010, in Phased Array Systems and Technology
  (ARRAY), 2010 IEEE International Symposium on, 640--647

\bibitem[{van Weeren {et~al.}(2016)van Weeren, Williams, Hardcastle,
  {et~al.}}]{vanweeren-factor-2016}
van Weeren, R.~J., Williams, W.~L., Hardcastle, M.~J., {et~al.} 2016, ApJSS,
  223, 2

\bibitem[{Wayth {et~al.}(2015)Wayth, Lenc, Bell, {et~al.}}]{wayth-2015-gleam}
Wayth, R.~B., Lenc, E., Bell, M.~E., {et~al.} 2015, PASA, 32

\bibitem[{{Wells} {et~al.}(1981){Wells}, {Greisen}, \&
  {Harten}}]{1982-wells-fitsformat}
{Wells}, D.~C., {Greisen}, E.~W., \& {Harten}, R.~H. 1981, A\&AS, 44, 363

\bibitem[{Wilson {et~al.}(2011)Wilson, Ferris, Axtens, Brown, Davis, Hampson,
  Leach, Roberts, Saunders, Koribalski, Caswell, Lenc, Stevens, Voronkov,
  Wieringa, Brooks, Edwards, Ekers, Emonts, Hindson, Johnston, Maddison,
  Mahony, Malu, Massardi, Mao, McConnell, Norris, Schnitzeler, Subrahmanyan,
  Urquhart, Thompson, \& Wark}]{atca-broadband-backend-2011}
Wilson, W.~E., Ferris, R.~H., Axtens, P., {et~al.} 2011, MNRAS, 416, 832

\bibitem[{Ziv \& Lempel(1977)}]{ziv-1977-lz77}
Ziv, J. \& Lempel, A. 1977, IEEE Transactions on Information Theory, 23, 337

\end{thebibliography}

\end{document}